\documentclass{aa}
\usepackage{graphicx}

\begin{document}

\title{The Mass Function of the Arches Cluster from Gemini Adaptive Optics Data\thanks{
Based on observations obtained with the Gemini North Telescope and the NASA/ESA Hubble Space Telescope.}}
   \author{A. Stolte
          \inst{1}\inst{,2}\and
          E. K. Grebel\inst{1}\and
          W. Brandner\inst{2}\and
          D. F. Figer\inst{3} 
          }

   \offprints{A. Stolte, \email{stolte@mpia-hd.mpg.de}}

   \institute{Max-Planck-Institut f\"ur Astronomie, K\"onigstuhl 17,
              69117 Heidelberg, Germany\\
              \email{stolte@mpia-hd.mpg.de, grebel@mpia-hd.mpg.de}
         \and
             European Southern Observatory, Karl-Schwarzschild-Str. 2, D-85748 Garching, Germany\\
             \email{wbrandne@eso.org}
         \and
             Space Telescope Science Institute, 3700 San Martin Drive, Baltimore, MD 21218, USA\\
             \email{figer@stsci.edu}
             }

\authorrunning{A. Stolte et al.}
\titlerunning{Mass Function of the Arches Cluster}

\date{Received 00 00 00 / Accepted 00 00 00}

\abstract{
We have analysed high resolution adaptive optics (AO) science demonstration data 
of the young, massive stellar cluster Arches near the Galactic Center, 
obtained with the Gemini North telescope 
in combination with the University of Hawai'i AO system Hokupa'a.
The AO H and K' photometry is calibrated using HST/NICMOS observations
in the equivalent filters F160W and F205W obtained by Figer et al. (\cite{FKM}).
The calibration procedure allows a detailed comparison of the 
ground-based adaptive optics observations against
diffraction limited space-based photometry. The spatial resolution as 
well as the overall signal-to-noise ratio of the Gemini/Hokupa'a data
is comparable to the HST/NICMOS data.
The low Strehl ratio of only a few percent is 
the dominant limiting factor in the Gemini AO 
science demonstration data as opposed to space-based observations. After a thorough
technical comparison, the Gemini and HST data are used in combination to 
study the spatial distribution of stellar masses in the Arches cluster.
Arches is one of the densest young clusters known in the Milky Way, with
a central density of $\sim 3 \cdot 10^5\,M_\odot\,{\rm pc^{-3}}$ and a total
mass of about $10^4\,M_\odot$. 
A strong colour gradient is observed over the cluster field.
The visual extinction increases by $\Delta A_V \sim 10$ mag over a distance of 
15\arcsec\ from the cluster core.
Extinction maps reveal a low-extinction cavity in the densest parts of
Arches ($R \leq 5\arcsec$), indicating the depletion of dust due to stellar winds 
or photo-evaporation. We correct for the change in extinction over the field
and show that the slope of the mass function is strongly influenced by
the effects of differential extinction.
We obtain present-day mass function
slopes of $\Gamma \sim -0.8 \pm 0.2$ in the mass range $6\!<\!M\!<\!65\ M_\odot$
from both data sets. The spatial analysis reveals a steepening of 
the mass function slope from close to zero in the cluster center to
$\Gamma \sim -1.7 \pm 0.7$ at $R > 10\arcsec$, in accordance 
with a Salpeter slope ($\Gamma = -1.35$).
The bias in the mass function towards high-mass stars in the Arches
center is a strong indication for mass segregation. The dynamical and relaxation timescales
for Arches are estimated, and possible mass segregation effects are discussed
with respect to cluster formation models. 
\keywords{Open clusters and associations: individual: Arches -- Stars: luminosity function, mass function -- Stars: early-type -- Stars: formation -- ISM: dust, extinction -- Instrumentation: adaptive optics}
}

\maketitle

\section{Introduction}

The Galactic Center (GC) is the most extreme star forming environment 
within the Milky Way. High stellar and gas densities, turbulent motion,
tidal torques exerted by the steep gravitational potential, magnetic fields
and an intense radiation field determine the physical environment of star
formation in the GC region. 
Although disruptive forces exerted by the gravitational and radiation
fields counteract the agglomeration of material, the high gas and dust
densities cause star formation in the GC environment to be most efficient. 
In particular, the formation of high mass stars and massive clusters is 
more successful than in any other region of the Milky Way. 

A detailed study of star
formation processes and the stellar content of the GC region has until recently
been limited to the brightest and most massive stars due to the large amount of 
extinction ($A_V \sim 30$ mag) along the line of sight. 
Additional constraints are imposed due to the spatial resolution at the GC 
distance of $\sim 8$ kpc ($DM = 14.47 \pm 0.08$ mag, e.g., McNamara et al. 
\cite{McNamara2000}), 
much farther than nearby star forming regions such as the Orion or $\rho$ Ophiuchi 
star forming complexes, which have been studied in greater detail to date.
Only with the advent of deep, high resolution near-infrared instruments, the 
analysis of stellar populations in young star clusters near the GC has become feasible.

During the past few years, it has become evident that three out of four young
starburst clusters known in the Milky Way are located in the 
GC region - namely, the \object{Arches} and Quintuplet clusters, as well as the 
Galactic Center Cluster itself. With a cluster age of only a few Myr for 
Arches and Quintuplet, the question arises how many clusters do actually form 
in the densest environment of the Milky Way.
The 2MASS database yielded new insights into the estimated number of star clusters
hidden in the dense stellar background. Dutra \& Bica (\cite{DB2000}, \cite{DB2001}) 
report the detection of new cluster candidates
of various ages located in the innermost 200 pc of the Galaxy found in 2MASS. 
Numerical simulations by Portegies Zwart et al. (\cite{PZ2001}) suggest 
that clusters with properties similar to the massive Arches and Quintuplet 
may have formed in the past in the innermost 200 pc, but were then dispersed
and are now indistinguishable from the dense stellar background.
As dynamical evolution timescales are short due to the strong tidal field 
in the GC region (Kim et al. \cite{Kim1999}),
young star clusters are disrupted quickly after formation, contributing to the 
Galactic bulge population. Thus, only the youngest clusters remain intact for 
the study of star formation in this extraordinary environment.

The Arches cluster, at a projected distance of only 25 pc from the GC
(assuming a heliocentric distance of 8 kpc to the GC),
is one of the most massive young clusters known in the Milky Way. 
With an estimated mass of about $10^4\,M_{\sun}$ and a central density
of $3 \cdot 10^5\,M_{\sun}\,{\rm pc}^{-3}$, Arches is the densest young star 
cluster (YC) known (Figer et al. \cite{FKM}).
From physical properties of Wolf-Rayet stars, the age of the cluster is estimated 
to be between 2 and 4.5 Myr (Blum et al. \cite{Blum2001}). 
The stellar content of Arches has been 
studied by Figer et al. (\cite{FKM}) using HST/NICMOS data. 
They derived a shallow initial
mass function in the range $6\!<\!M\!<\!120\,M_{\sun}$ with a slope of 
$\Gamma = -0.7 \pm 0.1$, but with significant flattening observed in the innermost
part of the cluster ($\Gamma = -0.1 \pm 0.2$). 

Most young star clusters and associations in the Milky Way display a mass 
function close to a Salpeter (1955) power law with a slope of $\Gamma = -1.35$.
Several such star forming regions have been studied by Massey et al. 
(\cite{Massey1995a}), yielding slopes in the range $-0.7 < \Gamma < -1.7$
with an average of $-1.1 \pm 0.1$, which leads these authors to conclude 
that within the statistical limits no deviation from a Salpeter slope is observed.

A flat mass function as observed in Arches implies an overpopulation of 
the high-mass end as compared to ``normal'' clusters. 
The special physical conditions in the GC region have been suggested 
to enhance the formation of massive stars, thereby resulting in a flattened 
mass function (Morris \cite{Morris1993}).
The formation of high-mass stars in itself poses serious problems for 
the standard core collapse and subsequent accretion model, as radiation pressure 
from the growing star is capable of reversing the gas infall as soon as
the mass is in excess of $10\,M_{\sun}$ (Yorke \& Kr\"ugel \cite{Yorke1977}). 
Assuming disk accretion instead of spherical infall, the limiting mass may be
increased to $15\,M_{\sun}$ (Behrend \& Maeder \cite{Behrend2001}), 
still far below the mass
observed in O-type stars. 
Various scenarios are suggested to circumvent this problem. 
Simulations with enhanced accretion rates and collision probabilities
in dense cluster centers (Bonnell et al. \cite{Bonnell1998a}), 
as well as growing accretion rates depending on the mass of 
the accreting protostar (Behrend \& Maeder \cite{Behrend2001}), allow stars of up to 
$100\,M_{\sun}$ to form in the densest regions of a rich star cluster.
In case of the GC environment,
a higher gas density may lead to a higher accretion rate and/or to a
longer accretion process in the protostellar phase. 
As long as the gravitational potential is strongly influenced by 
the amount of gas associated with the cluster, gas infall causes 
a decrease in cluster radius and subsequent increase in the collision rate,
reinforcing the formation of high-mass stars.
Physical processes such as gravitational 
collapse or cloud collisions scale with the square root of the local density,
$\sqrt{\rho}$ (Elmegreen \cite{Elmegreen1999}, \cite{Elmegreen2001}), 
causing an enhanced star formation rate
(SFR) in high density environments. Elmegreen (\cite{Elmegreen2001}) shows that the total 
mass as well as the maximum stellar mass in a cluster strongly depends on the
SFR and local density. This is confirmed by observations of high-mass stars
found predominantly in the largest star forming clouds (Larson \cite{Larson1982}). 

Both the growing accretion and the collision scenario predict the high-mass 
stars to form in the densest central region of a cluster, leading to 
primordial mass segregation, which may be evidenced in a flat mass 
function in the dense cluster center. As an additional physical constraint, 
both scenarios require 
the lower-mass stars to form first, and the highest-mass stars last in the
cluster evolution process. 
As the strong UV-radiation field originating from
hydrogen ignition in high-mass stars expells the 
remaining gas from the cluster center, the accretion process should be 
halted immediately after high-mass star formation.

The short dynamical timescales of compact clusters are, however, influencing the 
spatial distribution of stellar masses as well. On the one hand, high-mass stars
are dragged into the cluster center due to the gravitational potential
of the young cluster.
On the other hand, low-mass stars may easily be flung out of the cluster 
due to interaction processes, especially given star densities as high as in the
Arches cluster. The result of these processes would also be a flat mass
function in the cluster center, steepening as one progresses outwards
due to dynamical mass segregation. 
Dynamical segregation is predicted to occur within one relaxation time
(Bonnell \& Davies \cite{Bonnell1998b}), which for compact clusters is 
only one to a few Myr, and should thus be well observable in Arches 
in the form of a spatially varying mass function. 

In addition to the internal segregation process, the external GC tidal field
exerts shear forces tearing apart the cluster entity.
N-body simulations by Kim et al. (\cite{Kim2000}) yield tidal disruption timescales 
as short as 10 to 20 Myr in the GC tidal field. 
We expect to find a mixture of all these effects in the Arches cluster.

We have analysed adaptive optics (AO) data obtained under excellent seeing conditions
with the Gemini North 8m telescope in combination with the 
University of Hawai'i (UH) AO system Hokupa'a. We are investigating  the presence of
radial variations in the mass distribution within the Arches cluster. 
We compare our ground-based results in detail to the HST/NICMOS data presented 
in Figer et al. (\cite{FKM}, hereafter FKM), discussing possible achievements and 
limitations of ground-based, high-resolution adaptive optics  versus space-based 
deep NIR photometry. 

In Section 2, we will introduce the data and describe the reduction and 
calibration processes. In this context, a thorough investigation of the 
quality of ground-based adaptive optics photometry as compared to 
space-based diffraction limited observations will be presented.
The photometric results derived from colour-magnitude diagrams and
extinction maps will be discussed in Section 3.
A comparison of Gemini and HST luminosity functions will be given in 
Section 4. The mass functions will be derived in Section 5,
and their spatial variation will be discussed with respect to 
cluster formation scenarios. We will estimate the relevant timescales
for cluster evolution for the Arches cluster in Section 6, and 
discuss the implication on the dynamical evolution of Arches.
We will summarise our results in Section 7.

\section{Observations and Data Reduction} 
\label{obssec}

\subsection{Gemini/Hokupa'a Data}
\label{datasec}

We analysed $H$ and $K^\prime$ images of the Arches cluster center
obtained in the course of the Gemini science demonstration
at the Gemini North 8m telescope located on Mauna Kea, Hawai'i, 
at an altitude of 4200 m above sea level. \\
Gemini is an alt-azimuth-mounted telescope with a monolithic primary mirror
and small secondary mirror optimised for IR observations.
The telescope is always used in Cassegrain configuration with 
instruments occupying either the upward looking Cassegrain port
or one of three sideward facing ports. 
The University of Hawai'i adaptive optics (AO) system Hokupa'a 
is a 35 element curvature sensing AO system (Graves et al. \cite{Graves2000}),
which typically delivers Strehl ratios between 5 \% and 25 \% in the $K$-band.

Hokupa'a is operated with the near-infrared camera QUIRC 
(Hodapp et al. \cite{Hodapp1996}), equipped with
a $1024 \times 1024$ pixel HgCdTe array. The plate scale is 
19.98 milliarcseconds per pixel, yielding a field of view (FOV) of 20.2\arcsec. 
The images are shown in Fig. \ref{images}, along with the HST/NICMOS images
used for calibration.

\begin{figure*}
\vspace*{4ex}
{\bf Fig. 1 is available in jpeg format on astro-ph} \\
\\
\caption{Upper panels: Gemini/Hokupa'a $H$ (left) and $K^\prime$ (right) images, 
20\arcsec $\times$ 20\arcsec (0.8 pc $\times$ 0.8 pc) 
with a spatial resolution of less than 0.2\arcsec. North is up and East is to the right.
Lower panels: HST/NICMOS F160W and F205W images from Figer et al. (\cite{FKM}), geometrically
transformed to the Gemini FOV.}
\label{images}
\end{figure*}

The observations were carried out between July 3 and 30, 2000.
12 individual 60 s exposures, dithered in a 4 position pattern 
with an offset of 16 pixels (0.32\arcsec) between subsequent frames, 
were coadded to an $H$-band image
with a total integration time of 720 s. In $K^\prime$, the
set of 34 dithered 30 s exposures obtained under the best observing conditions
was coadded to yield a total exposure time of 1020 s. 
The full width at half maximum (FWHM) of the point spread function (PSF) 
was 9.5 pixels (0.19\arcsec) in $H$ and 6.8 pixels (0.135\arcsec)
in $K^\prime$. The observations were carried out at an airmass of 1.5,
the lowest airmass at which the Galactic Center can be observed from Hawai'i.

The $H$-band data were oversampled, and $2\times 2$ binning was
applied to improve the effective signal-to-noise ratio per resolution element,
allowing a more precise PSF fit.
A combination of long and short exposures has been used to 
increase the dynamic range.
For the short exposures, 3 frames with 1\,s exposure time have been coadded 
in $H$, and 16 such frames in $K^\prime$.
See Table \ref{obstable} for the observational details.
In the long exposures, the limiting magnitudes were about 21 mag in $H$ and 
20 mag in $K^\prime$. Note that the completeness limit in the crowded regions 
was significantly lower. The procedure used for completeness correction
will be described in detail in section \ref{incsec}.
\\
\begin{table*}
\centering
\caption{Gemini/Hokupa'a and HST/NICMOS observations}
\label{obstable}
\begin{tabular}{lcrrrlccc}
\hline
\noalign{\medskip}
Date & Filter & single exp. & ${\rm n_{exp}}$& exp. total & det. limit &\
$\sigma_{\rm back}$ &resolution & diffraction limit \\
\noalign{\medskip}
\hline
\noalign{\medskip}
\multicolumn{9}{c}{Gemini} \\
\noalign{\medskip}
\hline
\noalign{\smallskip}
05/07/2000 & H & 1 s & 3 &   3 s  & 18.5 mag & 7.74 & 0.17\arcsec & 0.05\arcsec\\
05/07/2000 & H & 60 s & 12 & 720 s & 21 \ \ mag & 0.19 & 0.20\arcsec & 0.05\arcsec\\
30/07/2000 & K$^\prime$ & 1 s & 16 & 16 s & 17.5 mag & 3.73 & 0.12\arcsec & 0.07\arcsec\\        
09/07/2000 & K$^\prime$ & 30 s & 34 & 1020 s & 20 \ \ mag & 0.22 & 0.13\arcsec & 0.07\arcsec\\
\noalign{\smallskip}
\hline
\noalign{\medskip}
\multicolumn{9}{c}{HST} \\
\noalign{\medskip}
\hline
\noalign{\medskip}
14/09/1997 & F160W & & & 256 s & 21 mag & 0.04 & 0.18\arcsec & 0.17\arcsec\\
14/09/1997 & F205W & & & 256 s & 20 mag & 0.15 & 0.22\arcsec & 0.21\arcsec\\
\noalign{\smallskip}
\hline
\end{tabular}
\end{table*}

\subsubsection{Data reduction}
\label{redsec}

The data reduction was carried out by the Gemini data reduction team, F. Rigaut, 
T. Davidge, R. Blum, and A. Cotera. The procedure as outlined in the 
science demonstration report\footnote{The description of the Galactic Center dataset can be \newline found at http://www.gemini.edu/gallery/observing/ \newline release\_doc/manual.html, and the Gemini North science \newline demonstration data are publicly available at http://www.gemini.edu/sciops/data/dataSV.html.}
was as follows: Sky images, obtained after the short observation period 
when the Galactic Center was in culmination, were averaged using median clipping 
for star rejection, and then subtracted from the individual images. 
The frames were then flatfielded and corrected for bad pixels and cosmic ray
hits. After inspecting the individual frames with respect to signal-to-noise ratio
and resolution, and background adjustment, the images with sufficient quality were
combined using sigma clipped averaging. The final images were scaled to counts per 
second. For the analysis presented in this paper, this set
of images reduced by the Gemini reduction team has been used.

\subsubsection{Photometry}
\label{photsec}

The photometry was performed using the IRAF\footnote{IRAF is distributed by the National Optical Astronomy Observatories, which are operated by the Association of Universities for Research in Astronomy, Inc., under cooperative agreement with the National Science Foundation.}
(Tody \cite{Tody1993}) DAOPHOT implementation 
(Stetson \cite{Stetson1987}).
Due to the wavelength dependence of the adaptive optics correction and 
anisoplanatism\footnote{The anisoplanatism describes the degradation of
the AO correction due to differences between wave fronts coming from different
directions.} 
over the field, the $H$ and $K^\prime$ data
have been treated differently for PSF fitting. While in $H$ the PSF radius 
increases significantly with distance from the guide star, 
with a radially varying FWHM in the range of 0.18\arcsec\ to 0.23\arcsec\ 
(see the science demonstration data description), 
the $K^\prime$ PSF was nearly constant over the field
(0.125\arcsec\ to 0.135\arcsec). This behaviour is expected from an AO system,
as the isoplanatic angle\footnote{The isoplanatic angle is defined as the 
angular distance for which the rms error of the wave front is less than 1 radian.}
$\theta_0$ varies as $\lambda^{6/5}$, 
yielding a 1.4 times larger $\theta_0$ in $K^\prime$ than in $H$, 
resulting in a more uniform PSF in $K^\prime$ across the field of view. 

As obscuration due to extinction decreases with increasing wavelength,
many more stars are detected in $K^\prime$ than in $H$.
For comparison, the number of objects found with $K^\prime < 20$ mag and
uncertainty $\sigma_{K^\prime} < 0.2$ mag was 1017 (1020 s effective exposure time), 
while for $H < 20$ mag and $\sigma_H < 0.2$ mag
we detected only 391 objects (720 s effective exposure time), 
where in both cases visual inspection led to the conclusion
that objects with photometric uncertainties below 0.2 mag were real detections.
On the other hand, the increased stellar number density in $K^\prime$ leads to
increased crowding effects,
such that we decided to use a non-variable PSF for the $K^\prime$-band data
after thorough investigation of the results of a quadratically, linearly or
non-varying PSF. It turns out that, due to a lack of isolated stars for 
the determination of the PSF variation across the field, the mean uncertainty is 
{\sl lower} and the number of outliers with unacceptably large uncertainties reduced 
when a non-variable PSF is used. 
Thus, the 5 most isolated stars on the $K^\prime$ image, which were
well spread out over the field, were used to derive the median averaged PSF 
of the long exposure.
In the case of the short exposure, where due to the very short integration 
time faint stars are indistinguishable from the background, 
leaving more `uncrowded' stars to derive the shape
of the PSF, 7 isolated stars could be used. 
In $K^\prime$, the best fitting function was an elliptical 
Moffat-function with $\beta = 2.5$.

Due to the lower detection rate on the $H$-band image crowding is less severe,
while the PSF exhibits more pronounced spatial variations than in $K^\prime$.
We thus used the quadratically variable option of the DAOPHOT {\sl psf} and
{\sl allstar} tasks for our $H$-band images, with 27 stars to determine a 
median averaged PSF function and residuals. 
The best fitting function was a Lorentz function
on the binned $H$-band image.
In both filters, the average FWHM of the data has been used as the PSF 
fitting radius, i.e., the kernel
of the best-fitting PSF function, to derive PSF magnitudes of the stars. 

The short exposures have been used 
to obtain photometry of the brightest stars, which are saturated in the 
long exposures. The photometry of the long and short exposures 
agreed well after atmospheric extinction correction in the form of a constant 
offset had been applied.
The saturation limit was 13.0 mag in $H$ and 13.3 mag in $K^\prime$.
At fainter magnitudes, the photometry of both images 
was indiscernible within the uncertainties for the 
bright stars, and the better quality long exposure values were used.
Furthermore, a comparison of the bright star photometries was used to 
estimate photometric uncertainties (see Sect. \ref{incsec} and Table \ref{uncertab}).
\\
\begin{figure*}
\includegraphics[width=8cm,angle=270,clip=]{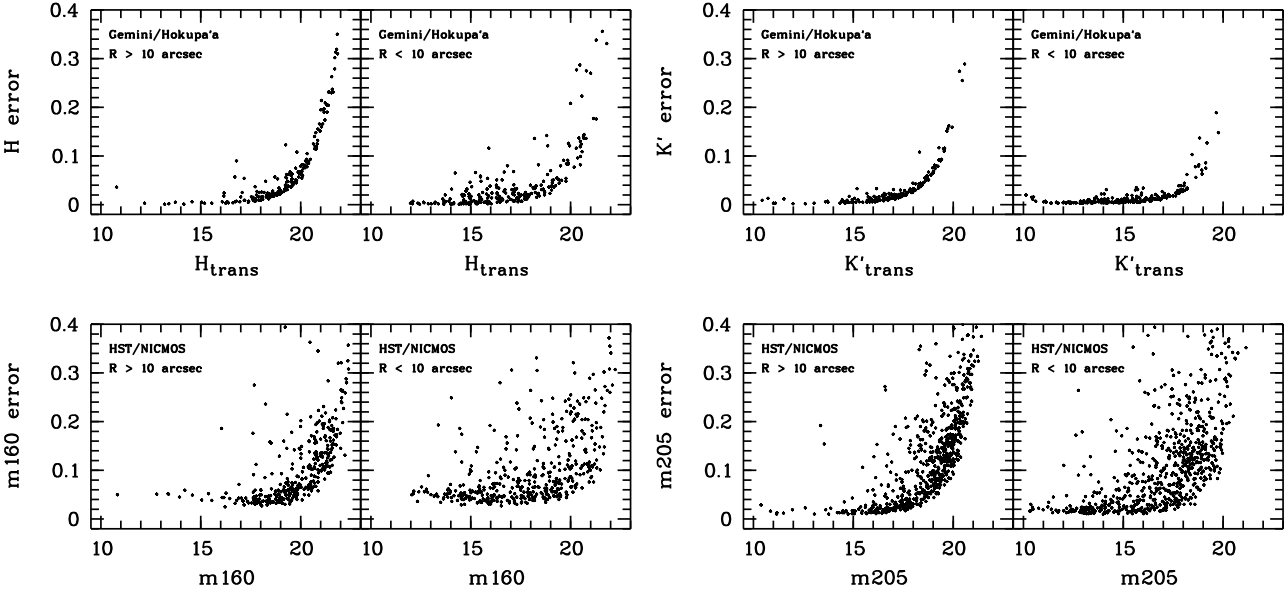}
\caption{Formal DAOPHOT photometric uncertainties. Top panel: Gemini/Hokupa'a $H$ and 
$K^\prime$ photometry transformed to HST/NICMOS F160W and F205W filter magnitudes,
$m160$ and $m205$.
Bottom panel: HST/NICMOS $m160$, $m205$ photometry of the same $20\times20\arcsec$ 
field (shown in Fig. \ref{images}).}
\label{magerr}
\end{figure*}

\subsubsection{Photometric Calibration}
\label{calsec}

To transform instrumental into apparent magnitudes, we used
the HST/NICMOS photometry of Figer et al. (\cite{FKM}) as local standards. 
The advantage of this procedure lies
in the possibility to correct for remaining PSF deviations over the field,
e.g., due to a change in the Strehl ratio with distance from the guide star
or due to the increased background from bright star halos in the cluster center.
Indeed, as will be discussed below, the spatial distribution of 
photometric residuals shows a mixture of these effects. 

\begin{figure*}
\includegraphics[width=8cm,angle=0,clip=]{arch.fig3a.ps} \hspace{1cm}
\includegraphics[width=8cm,angle=0,clip=]{arch.fig3b.ps}\\
\includegraphics[width=8.2cm,angle=270,clip=]{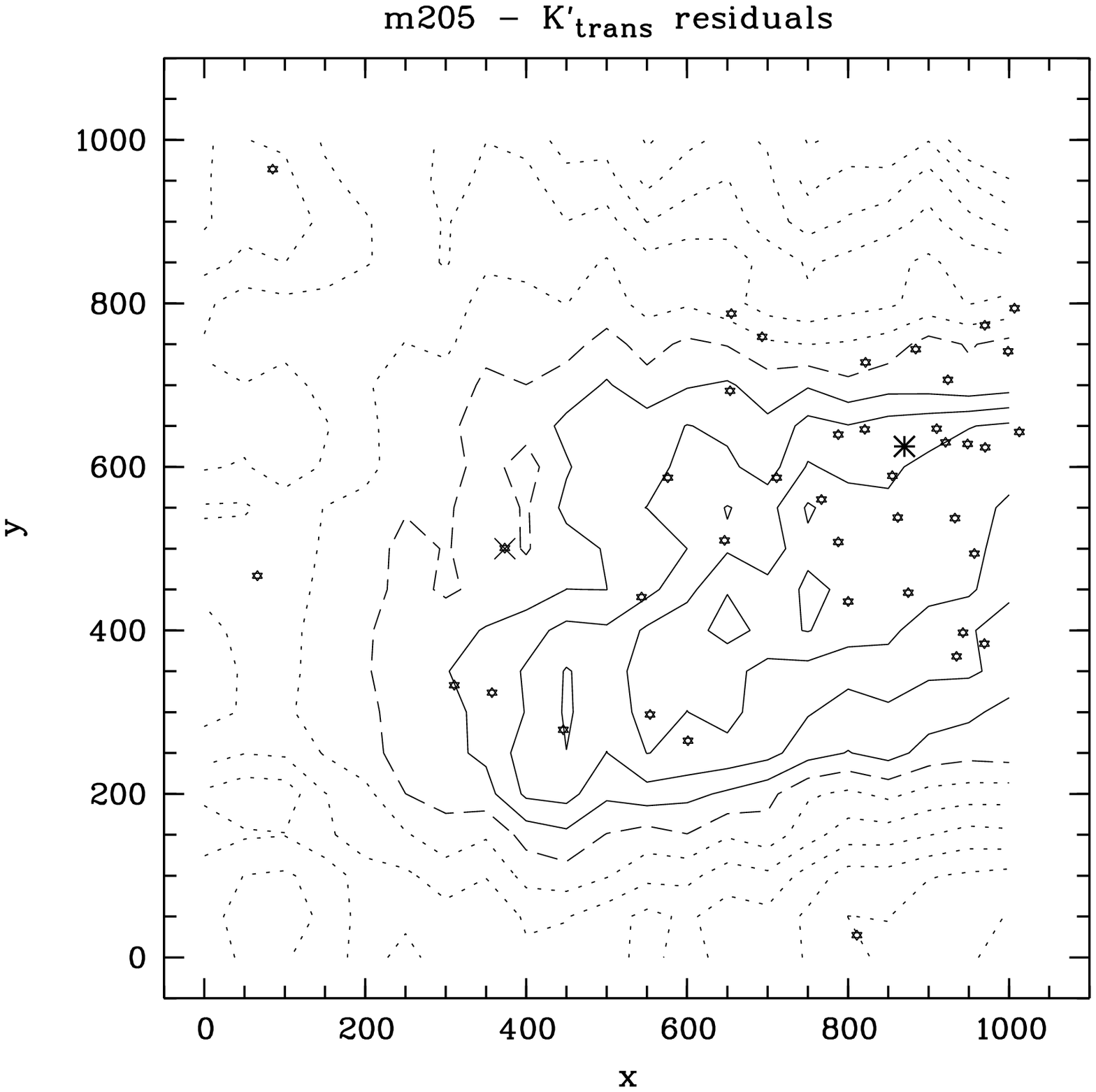} \hspace{0.7cm}
\includegraphics[width=8.2cm,angle=270,clip=]{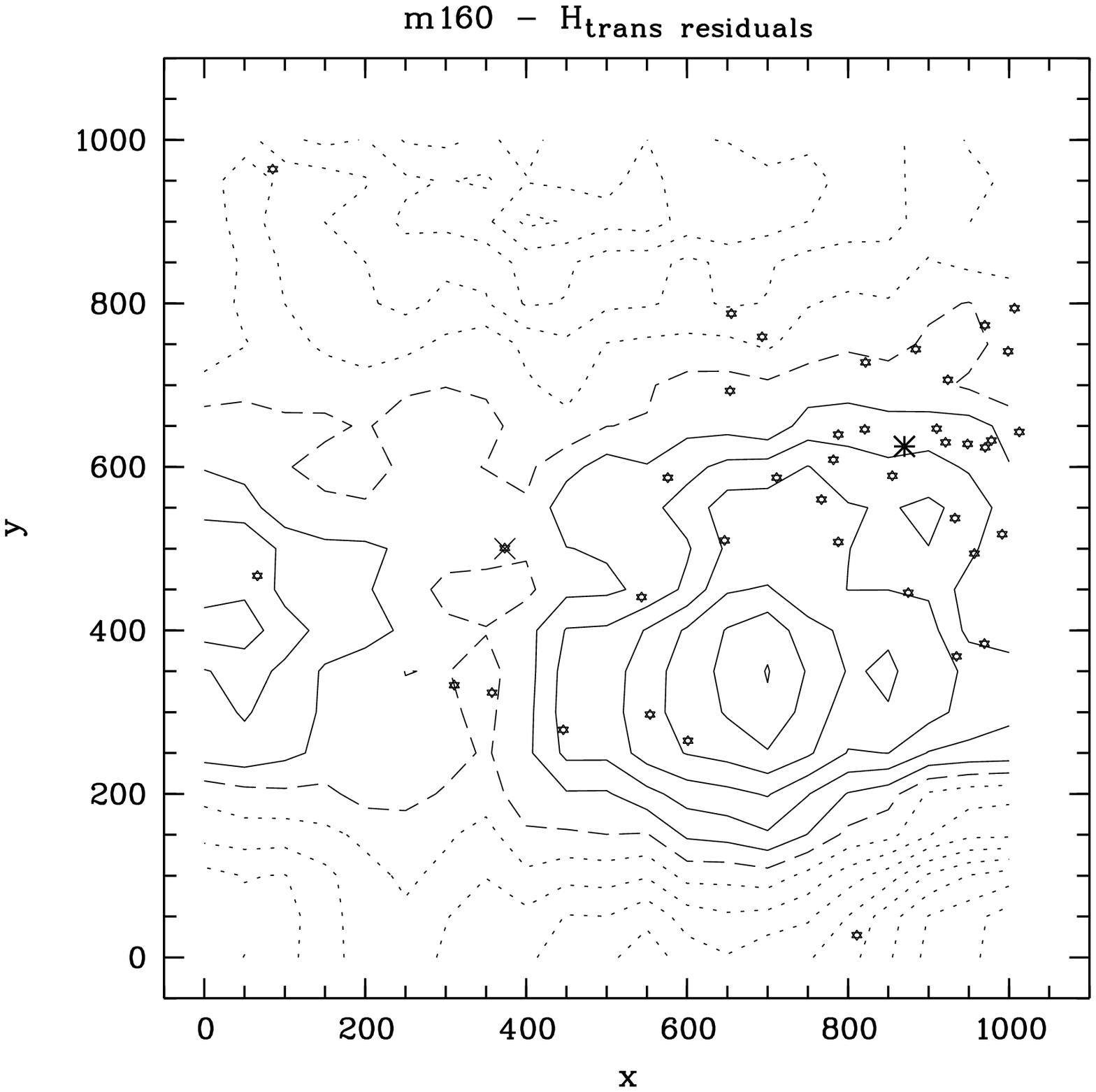}\\
\caption{Map of residuals of NICMOS vs. Gemini photometry (orientation as in 
Fig. \ref{images}). \newline
Left panels: $m205 - K^\prime_{\rm trans}$, right panels: $m160 - H_{\rm trans}$ \newline
Stars have been binned into areas of $50 \times 50$ pixels, and the value displayed 
shows the average of all stars in each bin. Statistical fluctuations are large
due to the varying number of stars in each pixel intervall, but the overall 
trends are clearly visible. The position of the guide star is marked by a cross,
the cluster center as determined from the HST/NICMOS F205W image as an asterisk,
and stars denote stars resolved in the Gemini images with magnitudes brighter than
$K^\prime_{\rm trans}=13$ mag. The strong correlation between the position of bright
stars and positive residuals reveals the tendency of PSF fitting photometry
applied to AO data to overestimate the flux of bright sources, and underestimate
the flux of faint sources in their vicinity (see text for discussion).}
\label{kresmap} 
\end{figure*}

We were able to use approximately 380 stars to derive colour equations.
The residuals obtained for these stars after calibration allow a detailed
analysis and correction of field variations.
The colour equations to transform Gemini instrumental $H$ and $K^\prime$
magnitudes to magnitudes in the HST/NICMOS filter system were determined 
using the IRAF PHOTCAL package, yielding:
\vbox{
\begin{displaymath}
m160 = H_{\rm inst} + 0.001 (\pm 0.017) \cdot (H - K^\prime)_{\rm inst}
\end{displaymath}
\begin{displaymath}
\hspace*{5cm} - 0.028 (\pm 0.031)\ {\rm mag}
\end{displaymath}
\begin{displaymath}
m205 = K^\prime_{\rm inst} + 0.023 (\pm 0.008) \cdot (H - K^\prime)_{\rm inst}
\end{displaymath}
\begin{displaymath}
\hspace*{5cm} - 1.481 (\pm 0.016)\ {\rm mag},
\end{displaymath}
}
where $H_{\rm inst}$ and $K^\prime_{\rm inst}$ are the Gemini instrumental 
magnitudes, and $m160$ and $m205$ correspond to magnitudes obtained with
the NICMOS broadband filters F160W and F205W, respectively.
After the transformation had been applied, it turned out that the residual 
magnitudes in the two independent fitting parameters $H - K^\prime$
and $K^\prime$ still varied systematically over the field.
As can be seen in Fig. \ref{kresmap}, the variation is {\sl not} a simple radial 
variation increasing with distance from the guide star (GS), but a mixture
of displacement from the GS position (shown in the contour map in Fig. 
\ref{kresmap} as a cross), and the position with respect to the cluster center
or bright stars in the field.
Fortunately, the variation was well behaved in the $Y$-direction, and 
could be fitted by a fourth order polynomial. Close inspection showed 
that two areas on the QUIRC array showed a remaining photometric offset compared 
to the HST photometry. In the region $400 < X < 700$ pixels and $Y < 250$ pixels, 
the magnitude was underestimated by $0.1$ mag. For $X > 850$ pixels 
and $Y > 850$ pixels, i.e., the upper right corner, the $K^\prime$ magnitude was 
overestimated by $0.25$ mag (however, note that there are only 7 stars 
in this region). For a discussion of these effects, see Sect. \ref{ressec}.
We corrected for these offsets before deriving the $Y$-correction, which then 
showed remarkable homogeneity over the entire field.
This smooth correction function is probably due to discrepancies in the 
dome flat field illumination versus sky exposures. 
The correction 
was then applied to transformed $K^\prime$ and $H-K^\prime$ magnitudes,
denoted $K^\prime_{\rm trans}$ and $(H-K^\prime)_{\rm trans}$ from now on.
The $H_{\rm trans}$ magnitude was calculated from the corrected $K^\prime_{\rm trans}$ and 
$(H - K^\prime)_{\rm trans}$ values.

An additional advantage of this procedure is the independence on uncertainties in 
colour transformations at large reddening and non-main sequence colours,
as opposed to colour transformations derived from typical main sequence
standard stars.
Using the HST photometry as local standards, we are naturally in equal
colour and temperature regimes,
allowing the direct comparison of the Gemini and HST photometry.
For most parts of the paper, we remain in the HST/NICMOS system.
We use the colour equations obtained in Brandner et al. (\cite{Brandner2001},
hereafter BGB) to transform typical main sequence colours 
and theoretical isochrone magnitudes into the HST/NICMOS system where 
indicated. This allows us to transform mainly unreddened main-sequence stars,
for which the BGB colour transformations have been established.
The only exceptions are the two-colour diagram (Sect. \ref{ccdsec}) 
and the derivation of the extinction variation from colour gradients 
(Sect. \ref{trendsec}), where the extinction law is needed to 
determine the reddening path. 
We will use the notation ``$m160$'' and ``$m205$'' as in FKM for magnitudes in the
HST/NICMOS filters, 
and ``$H_{\rm trans}$'' and ``$K^\prime_{\rm trans}$'' for the Gemini/Hokupa'a 
data calibrated to the NICMOS photometric system. 
HST magnitudes transformed to the ground-based 2MASS system will be denoted by
$(JHK_s)_{\rm 2MASS}$ or simply $JHK_s$.
\\

\subsubsection{Discussion of the residual maps}
\label{ressec}

The behaviour of the residual of the HST/NICMOS vs. Gemini magnitudes,
$m_{\rm NICMOS} - m_{\rm Gemini,trans}$, can be analysed in more detail when studying 
the residual map and the smoothed contour plot. In (Fig. \ref{kresmap}),
positive (negative) residuals correspond to overestimated (underestimated) flux.
From the map we denote a general tendency to overestimate the flux.
The contour maps show that positive residuals are correlated
with the position of bright stars on the $K^\prime$ image, both in the crowded
cluster center as well as in the area to the lower left, where a band of bright stars 
is located (see Fig. \ref{images}). This is the area where the $K^\prime$ magnitudes 
were found to be underestimated in Sect. \ref{calsec}, and thus the flux overestimated.
This suggests that the increased background
due to the uncompensated seeing halos of bright stars (cf. Sect. \ref{strehlsec}), 
inherent to AO observations, causes an overestimation of the flux of bright 
($K^\prime\stackrel{<}{\sim} 13$ mag)) sources. 
On the other hand, points with negative residuals are mainly correlated with fainter
stars ($K \stackrel{>}{\sim} 16$ mag), suggesting that this enhanced background
leads to an oversubtraction of the individually calculated 
background of nearby fainter stars. The result is an underestimate of the 
flux of companion stars in the vicinity of bright stars.
The correlation of positive residuals with the position of bright stars
seems to be less pronounced in the $H$-band image
(Fig. \ref{kresmap}). In $H$, the distance from the guide star is supposed to 
be more important due to the smaller size of the isoplanatic angle 
at shorter wavelengths and consequently more pronounced anisoplanatism.
Indeed, the smoothed residual contour plot shows a symmetry in the residuals
around the guide star, with close-to-zero residuals in the immediate vicinity
of the guide star, where the best adaptive optics correction can be achieved.
With increasing distance from the guide star, 
the residuals increase not only towards the bright
cluster, but also to the west (left in Fig. 1) of the field, 
indicating that remaining distortion 
effects from the AO correction are mixed with the problem of the proximity to 
bright stars as seen in $K^\prime$.

\subsubsection{Strehl ratio}
\label{strehlsec}

The Strehl ratio (SR) is defined to be the ratio of the observed peak-to-total flux
ratio
to the peak-to-total flux ratio of a perfect diffraction limited optical system.
This definition allows to compare the quality and photometric resolution of 
different optical systems using a single characteristic quantity. 
\begin{displaymath}
SR = (F_{\rm peak}/F_{\rm total})_{\rm obs} / (F_{\rm peak}/F_{\rm total})_{\rm theo}
\end{displaymath}
where $F_{\rm peak}$ is the maximum flux value of the PSF, and $F_{\rm total}$ is
the total flux, including the uncompensated halos induced by the natural seeing. 
The labels $obs$ and $theo$ refer 
to the observed and the theoretically expected flux, respectively.
Diffraction limited theoretical PSFs have been calculated using the 
{\sl imgen} task of the ESO data analysis package {\sl eclipse} (Devillard 
\cite{Devillard1997}).
The Strehl ratio in the Gemini $H$-band 720s exposure is found to be 2.5 \% compared
to 95 \% in the HST F160W image, and 7 \% in $K^\prime$ (1020s) 
compared to $\sim 90$ \% in F205W. The low SRs measured in the Gemini science
demonstration data indicate that more than 90 \% of the light of a star
is distributed into the resolution pattern of the AO PSF and the halo around
each star induced by the natural seeing. It has turned out that these
halos cause a significant limitation to the resolution and depth of the observations
in a crowded field, as faint stars can be lost to the enhanced background in the 
vicinity of bright objects. When comparing the HST and Gemini luminosity functions 
(Sect. \ref{lfsec}), this effect causes the main difference between both data sets.

In the case of very low Strehl ratios, the SR does not directly indicate the
fraction of the flux concentrated in the FWHM area of the PSF. A much larger 
fraction of the source flux can be used for PSF fitting in this case,
although the spatial resolution is limited by the large FWHM as compared to 
diffraction limited observations (cf. Tab. \ref{obstable}).
The ratio of the flux in the FWHM kernel of the compensated stellar image 
to the total flux, 
\begin{displaymath}
FR_{\rm obs} = F_{\rm FWHM}/F_{\rm total},
\end{displaymath}
may be determined by creating curves of growth for individual stars 
(Stetson \cite{Stetson1990}). 
The larger number of nearly isolated stars found on the $H$-band image
(used for PSF creation) 
allowed a reliable determination of $FR_{\rm obs}$ only in $H$, although
many stars on the $H$-band image were still too influenced by 
neighbours to study the aperture curve of growth in detail.
We were able to create well-behaved curves of growth for 7 stars.
As in the PSF fitting routine, we have used the FWHM of the PSF as kernel radius
and as the reference flux for the flux ratio determination.
These ratios range from 0.47 to 0.58, with an average of $0.53 \pm 0.05$,
i.e., $\sim$ 50 \% of the integrated point source flux are used for PSF fitting.
In addition, the variation of the flux ratio $FR_{\rm obs}$ over the field
serves as an indicator of the Strehl ratio variation. 
If the AO correction mechanism is the dominant factor determining the 
concentration of the flux into the FWHM kernel, the Strehl ratio and 
thus $FR_{\rm obs}$ should decrease with distance form the guide star, 
as the seeing correction worsens.
No correlation of the flux ratio with distance from the guide star is found.
Though these are small number statistics, this supports the suggestion 
that the sensitivity variations over the field are not predominantly due 
to increasing distance from the guide star (Sect. \ref{ressec}).

\subsubsection{Photometric uncertainties and incompleteness calculation}
\label{incsec}

\begin{figure}
\includegraphics[width=6cm,angle=270,clip=]{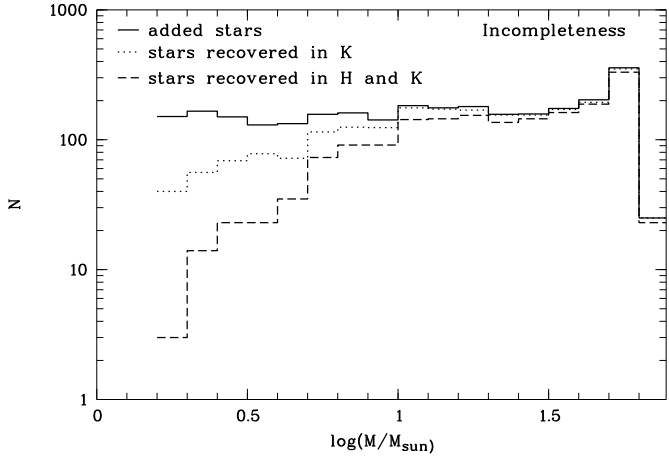}
\caption{Incompleteness tests performed on the Gemini $H$ and $K^\prime$ data.
The plot demonstrates the necessity to include the matching between single filter
observations to obtain a realistic estimate on the incompleteness, when matching
of objects is needed to create colour-magnitude- and two-colour-diagrams.}
\label{mfinc}
\end{figure}

For the incompleteness correction, artificial frames were
created with randomly positioned artificial stars. Magnitudes were
also assigned automatically in a random way. Due to the very crowded
field, only 40 stars were added to each artificial frame in order
to avoid significant changes in the stellar density. A total of 100 frames
was created for both the $H$ and $K^\prime$ deep exposures, leading
to a total of 4000 artificial stars used in the statistics.
In addition to the individual incompleteness in each band, the loss of
sources due to scatter of the main sequence generated by the more 
uncertain photometry
in the dense parts of the cluster was estimated. For this purpose the 
artificial $K^\prime$ stars were assigned a formal instrumental `colour'
of $(H - K^\prime) = 0.33$ mag (corresponding to 1.745 mag after photometric
transformation), 
derived from the average observed instrumental colour of the main sequence, 
and via this transformed into instrumental $H$ magnitudes. 
Artificial stars were inserted at the same positions in the $H$ and $K^\prime$
frames.
In this way the procedure also accounts for 
stars lost due to the matching of $H$ and $K^\prime$ data.
The artificial stars were calibrated
using the colour equations shown in Sect. \ref{calsec} , thus allowing us 
to estimate 
the loss of stars in the mass function derivation due to the applied main sequence 
colour selection (see Sect. \ref{mfsec}). 
This resulted in significantly larger corrections as compared
to the individual filter recoverage without matching and colour selection. 
As an example, the results for the mass function calculation performed
on the artificial stars are displayed in Fig. \ref{mfinc}.

As the recovery rate depends strongly on the stellar density and thus radial 
distance from the cluster center, the incompleteness correction will be determined
in dependence of the radial bin analysed when radial variations in the MF are 
studied (Sect. \ref{mfradsec}).

In addition to luminosity and mass function corrections, the artificial star tests
were used to estimate the real photometric uncertainties by comparing 
inserted to recovered magnitudes of the artificial stars. 
The median difference between the original and the recovered 
magnitude of the artificial stars, 
$\Delta m_{\rm arti} =\,<\!\!m_{\rm added} - m_{\rm DAOPHOT}\!\!>$, has
been used as an estimate of the real photometric uncertainty. To obtain the
median uncertainty, the intervalls $0 < \Delta m_{\rm arti} < 1$ and 
$-1 < \Delta m_{\rm arti} < 0$ have been treated individually, and the mean of the absolute
value of both median values, 
weighted with the number of objects in each intervall, is quoted in Tab. \ref{uncertab}. 
The overall flux deviation, including positive and negative deviations,
is close to zero for stars brighter than 20 mag ($< \pm 0.004$), and for fainter stars
becomes $-0.17$ and $-0.13$ in $H$ and $K^\prime$, respectively, showing a 
tendency to  underestimate the flux of faint stars. This tendency is more serious 
in the cluster center, where comparably large uncertainties are already observed for 
magnitudes fainter than 18 mag in both pathbands.

In a second test, 
photometry of the bright stars in the short exposures was compared to the
magnitudes of the deep exposures, yielding 
$\Delta m_{\rm sl} =\ <\!\!m_{\rm short} - m_{\rm long}\!\!>$
using the same procedure as for artificial stars. 
Note, however, that the quality of the short 
exposures is much worse than the one of the long exposures due to the high 
background noise. Therefore, the artificial star experiments, for which only
the deep exposures have been used, yield a more realistic estimate of the 
photometric uncertainties. The results of both tests are summarised in Table 
\ref{uncertab}.

The resulting uncertainty is roughly a factor of 2 to 3 larger than 
the theoretical magnitude uncertainty $\sigma_{\rm DAO}$ determined
from detector characteristics by DAOPHOT. The photometry in the cluster
center shows a larger uncertainty than the photometry in the outskirts.
As expected in a crowding-limited field, this also implies a reduced 
detection probability of faint sources in the center of the cluster.

\begin{table*}
\centering
\caption{Photometric uncertainties derived from artificial star experiments
($\Delta$m$_{\rm arti}$) and the comparison of short and long exposures 
($\Delta$m$_{\rm sl}$). The photometric uncertainties determined by DAOPHOT
are given for comparison. Note that for the brightest bin, 10 - 12 mag,
only the photometry of the short exposures was available. The higher 
starting bin in $H$ is due
to the calibration procedure of the inserted artificial stars. Magnitudes are 
given in the NICMOS system (F160W, F205W).}
\label{uncertab}
\begin{tabular}{ccccccccccc} 
\hline
\noalign{\smallskip}
Band & mag & $\Delta$m$_{\rm arti}$ & $\Delta$m$_{\rm sl}$ & $\sigma_{\rm DAO}$ & $\Delta$m$_{\rm arti}$ & $\Delta$m$_{\rm sl}$ & $\sigma_{\rm DAO}$ & $\Delta$m$_{\rm arti}$ & $\Delta$m$_{\rm sl}$ & $\sigma_{\rm DAO}$ \\
\noalign{\smallskip}
\hline
\noalign{\smallskip}
 & & \multicolumn{2}{c}{all} & \multicolumn{2}{c}{$R < 10$\arcsec} & \multicolumn{2}{c}{$R > 10$\arcsec} \\ 
\noalign{\smallskip}
\hline
\noalign{\smallskip}
H    & 12 - 14& 0.007 & 0.059 & 0.004 & 0.012 & 0.052 & 0.005 & 0.004 &   -   & 0.002 \\
     & 14 - 16& 0.015 & 0.061 & 0.008 & 0.025 & 0.065 & 0.010 & 0.011 & 0.035 & 0.004 \\
     & 16 - 18& 0.044 & 0.128 & 0.015 & 0.077 & 0.120 & 0.017 & 0.038 & 0.134 & 0.009 \\
     & 18 - 20& 0.119 &    -  & 0.036 & 0.167 &   -   & 0.039 & 0.098 &   -   & 0.030 \\
     &  $> 20$& 0.264 &    -  & 0.142 & 0.514 &   -   & 0.142 & 0.265 &   -   & 0.144 \\
\noalign{\medskip}
K$^\prime$ & 10 - 12 & 0.004 & - & 0.005 & 0.004 & -  & 0.005 & 0.003 &   -   & 0.006 \\
     & 12 - 14& 0.008 & 0.042 & 0.005 & 0.011 & 0.048 & 0.005 & 0.007 & 0.028 & 0.005 \\
     & 14 - 16& 0.027 & 0.072 & 0.007 & 0.041 & 0.086 & 0.007 & 0.020 & 0.042 & 0.006 \\
     & 16 - 18& 0.071 & 0.121 & 0.016 & 0.125 & 0.166 & 0.017 & 0.057 & 0.088 & 0.016 \\
     & 18 - 20& 0.206 &   -   & 0.060 & 0.280 &   -   & 0.073 & 0.189 &   -   & 0.056 \\
     &  $> 20$& 0.411 &   -   & 0.274 & 0.520 &   -   &    -  & 0.360 &   -   & 0.274 \\
\noalign{\smallskip}
\hline
\end{tabular}
\end{table*}

\subsection{HST/NICMOS data}
\label{nicdatasec}

HST/NICMOS observations have been obtained in the three broad-band filters
F110W, F160W and F205W, roughly equivalent to $J$, $H$ and $K$.
The basic parameters are included in Table \ref{obstable}.
For a detailed description of the HST data and their reduction 
see Figer et al. (\cite{FKM}).

\section{Photometric results}
\label{photressec}

\subsection{Radial colour gradient}
\label{trendsec}

\begin{figure}
\includegraphics[width=6cm,angle=270,clip=]{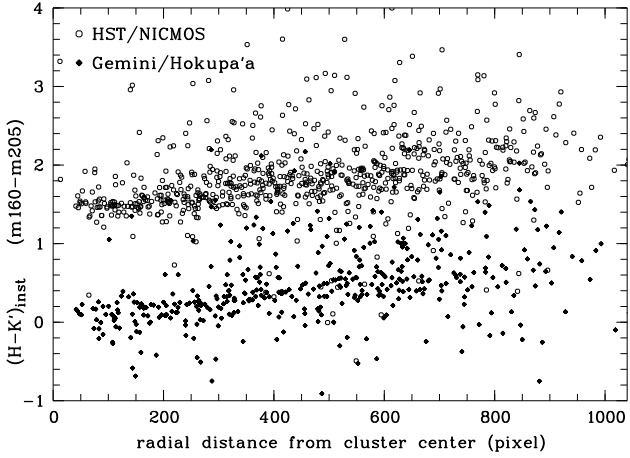}
\caption{Colour variation across the Gemini field, as observed in instrumental 
Gemini magnitudes and HST/NICMOS data. As a trend of increasing colour excess
with increasing distance from the Arches cluster center 
is observed in the two independent datasets, an instrumental effect as the cause 
for this variation is highly improbable.}
\label{hkinst}
\end{figure}

\begin{figure}
\includegraphics[width=8cm,clip=]{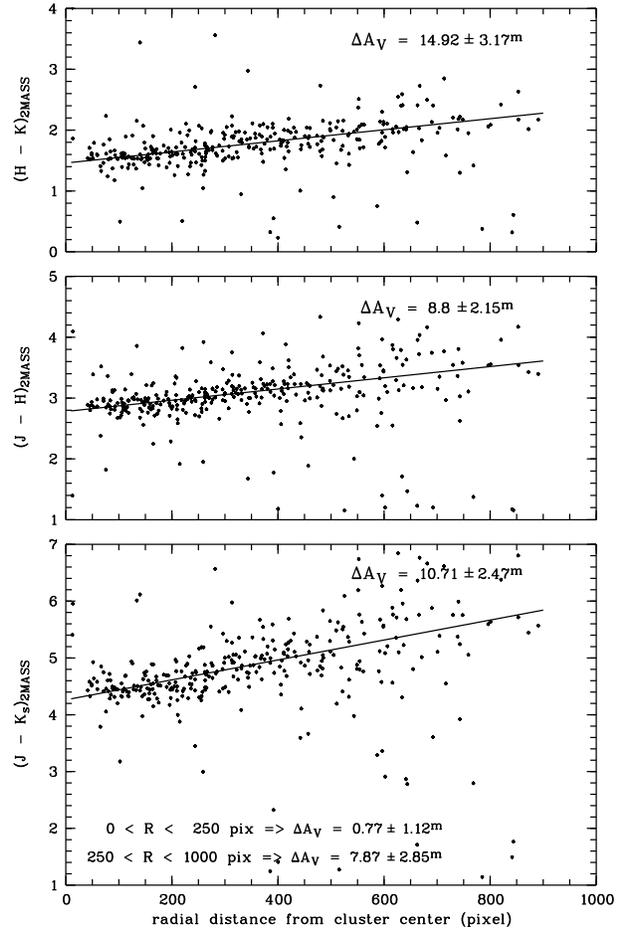}
\caption{Colour trends over the Arches field as observed in the HST/NICMOS data set within the area covered by the Gemini observations.
The radial distance from the cluster center is given in pixels on the Gemini
scale. For the calculation of $\Delta A_V$, the Rieke \& Lebofsky (\cite{RL85}) reddening law has 
been assumed.}
\label{hktrend}
\end{figure}

During the process of calibration we realised that a strong colour gradient is
present in the Arches field. 
In Fig. \ref{hkinst}, the $H-K^\prime$ colour is plotted against radial 
distance from the cluster center.
As the extinction law has not yet been derived
for HST filters, we have used the colour transformations in BGB
to transform NICMOS into 2MASS magnitudes. Though the 2MASS filters
deviate slightly from the standard Johnson $JHK$ filters used to determine the
extinction law (Rieke \& Lebofsky \cite{RL85}), we will be able to estimate 
the approximate amount of change in visual extinction across the field.
The extinction parameters from Rieke \& Lebofsky (\cite{RL85}) are given by
\begin{displaymath}
A_J/A_V = 0.282 \ \ \ A_H/A_V = 0.175 \ \ \ A_K/A_V = 0.112,
\end{displaymath}
where $A_{Filter}$ is the extinction in the given filter.
For a change in extinction of $\Delta A_V$ this leads to
\begin{eqnarray}
\Delta A_V = \Delta (A_J - A_H) / 0.107 \nonumber \\
\Delta A_V = \Delta (A_H - A_K) / 0.063 \nonumber \\
\Delta A_V = \Delta (A_J - A_K) / 0.170 \nonumber
\end{eqnarray}
We are able to measure the right-hand side of each expression 
from the transformed HST/NICMOS $JHK_s$ photometry. 
The resulting $\Delta A_V$ is given in each plot in Fig. \ref{hktrend}, 
together with the fitting uncertainty from the rms scatter in the colour. 

From the linear fits in Fig. \ref{hktrend} we see that
$A_V$ increases by about one order of magnitude 
over the entire field when moving outwards from 
the cluster center. The effect is
most pronounced in the HST $J-K_s$ vs. radius diagram (Fig. \ref{hktrend}, bottom), 
where the longest colour baseline is used. We derive a change in visual extinction 
of $\Delta A_V = 10.71 \pm 2.47$ mag over the Gemini field 
(1000 pixels $\hat{=}$ 0.8 pc).
Notably, if only the innermost 5\arcsec\ (250 pixels, 0.2 pc)
are fitted, no variation in $A_V$ is observed. 
When fitting the core separately, we get  
$\Delta A_V = 0.77 \pm 1.12$ mag for $R < 5\arcsec$ ($250$ pixels)
versus $\Delta A_V = 7.87 \pm 2.85$ mag for $5\arcsec < R < 20\arcsec$ 
($250 < R < 1000$ pixels). The latter value corresponds to 
$\Delta A_V = 10.5\ {\rm mag}/1000\ {\rm pix}^{-1}$, consistent 
with the trend over the entire field.
The small radial trend and low extinction value in the cluster center
indicates the local depletion of dust. This could be either due to
winds from massive stars or due to photo-evaporation of dust grains
caused by the intense UV-radiation field.

A change in $A_V$ of $\sim 10\ {\rm mag}/1000\ {\rm pix^{-1}}$ is also
consistent with the result found in  $J-H$,
$8.8\pm2.1\ {\rm mag}/1000\ {\rm pix^{-1}}$, 
while a larger value of $\Delta A_V = 14.9\pm 3.2$ mag is derived
from the $H-K_s$ plot. Due to the uncertainties we conclude that the 
extinction varies by $\Delta A_V \sim 9 - 15$ mag across the 
Arches field of 20\arcsec\ or 0.8 pc, most likely
closer to the lower value. This is in any case a tremendous change 
in the dust column density along the line of sight, with strong implications 
on limiting magnitudes and the potential detection of faint objects.

We have used a linear fit to the colour variation with radius for $R > 5\arcsec$ 
to correct for the strong change in reddening  observed in the outer cluster field.
The values for cluster center stars ($R < 5\arcsec$) have been left unchanged,
due to the large scatter and the very small trend found. Thus, these adjusted colours 
are scaled to the cluster center, where $A_V$ is lowest.
From the Rieke \& Lebofsky (1985) extinction law, the change in 
$K$-band magnitude with radius corresponding to the change in colour
can be derived as $\Delta A_K = 0.112/0.063 \cdot \Delta A_{H-K}$. 
We have used this relation to adjust the $K$-band magnitudes accordingly.
The 'dereddened' colour-magnitude diagrams will be shown in direct comparison with 
the observed CMDs in Sect. \ref{cmdsec} (cf. Fig. \ref{cmd}). 


\subsection{Extinction maps}
\label{extimapsec}
\begin{figure*}
\includegraphics[width=8cm,clip=]{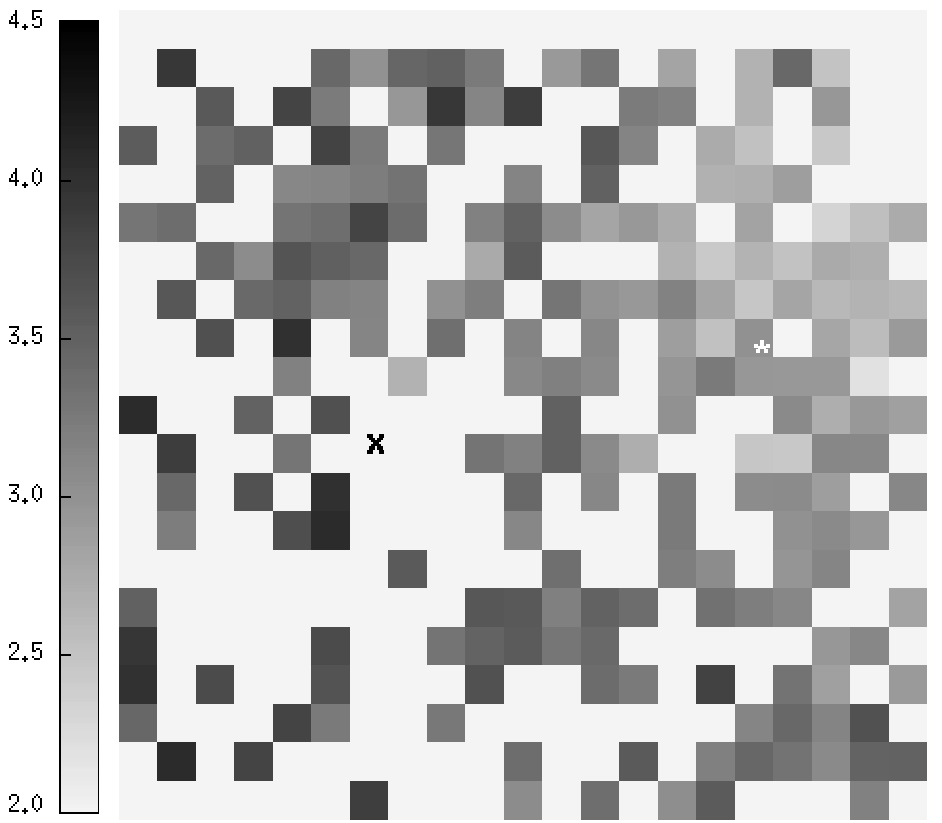} \hspace{1cm}
{\bf Fig. 7b is available in jpeg format on astro-ph} \\
\caption{$A_K$ extinction map, binned in the same manner as the residual map
in Fig. \ref{kresmap} (North is up, East is to the right). 
White spots are positions without stars for evaluation.
The individual extinction has been calculated by shifting the stars in the 
$K$ vs. $H - K$ colour-magnitude diagram to a 2 Myr isochrone offset bluewards
of the main sequence. Transformation to $A_K = 0$ mag has been performed afterwards,
to avoid large errors in the shifting procedure. 
This results in a minimum $A_K$ of 1.86 mag, and a maximum of 4.08 mag, 
assuming a Rieke \& Lebofsky (\cite{RL85}) extinction law.}
\label{extinct}
\end{figure*}

We can calculate the individual extinction for each object by shifting it along 
the reddening vector onto the isochrone. 
If we assume that intrinsic reddening plays only a minor role for most
of the stars within the cluster, we can combine the individual reddenings to 
estimate the spatial extinction variation.
In practice, we have applied a reddening of $A_V = 15$ mag to a 2 Myr
isochrone of the Geneva set of models (Lejeune \& Schaerer \cite{Lejeune2001}). 
This choice of reddening ensures that the isochrone serves as
a blue envelope for the bulk of the cluster stars, which are significantly
more reddened. 
The choice of the isochrone will be discussed in context with the mass function 
(Sect. \ref{mfsec}),
where the physical effects of the population model used are more important.
As discussed in Grebel et al. (\cite{G1996}), colours of main-sequence stars
also depend on binarity, stellar rotation, and intrinsic infrared excess. 
We ignore these effects here since we have no means to distinguish them 
from reddening effects.
In the worst case, we will overestimate the reddening for some stars.
We have then shifted the stars along the reddening path in colour-magnitude space,
again assuming that the relative slopes of the extinction law approximately 
follow a Rieke \& Lebofsky (\cite{RL85}) law. The resulting extinction map
for the Gemini photometry is shown in Fig. \ref{extinct}.

\begin{figure*}
\includegraphics[width=8cm,clip=]{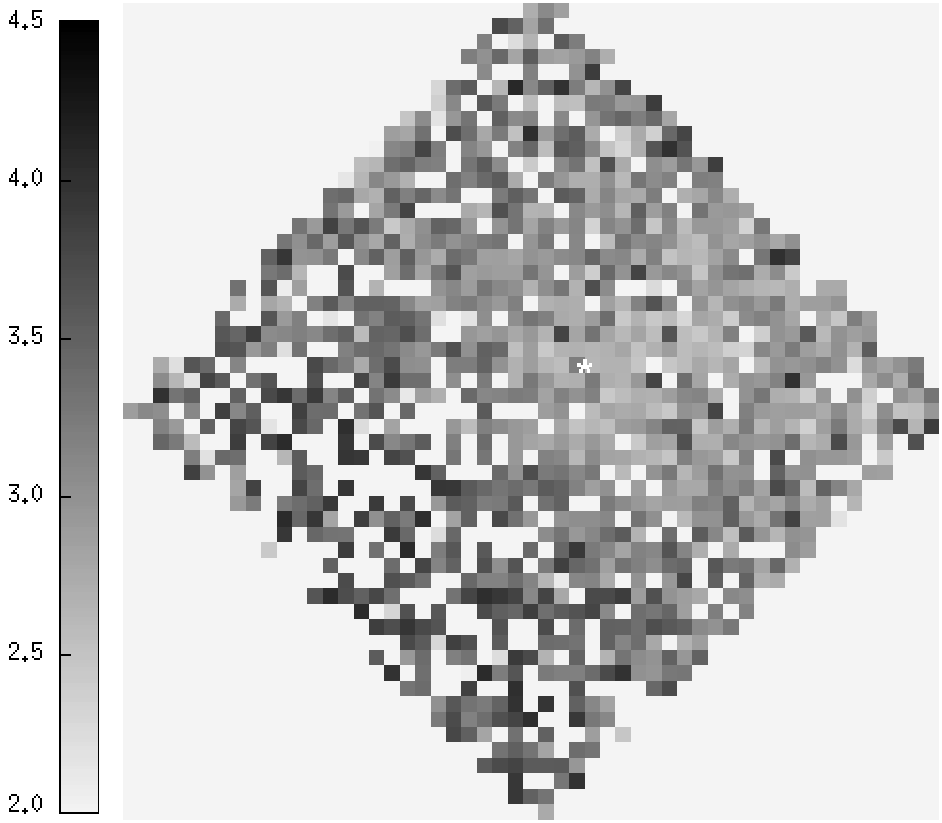} \hspace{1cm}
{\bf Fig. 8b is available in jpeg format on astro-ph}\\
\caption{$A_K$ extinction map derived from HST $m205$ photometry. See Fig. \ref{extinct}
for details. The coordinate transformed HST/NICMOS F205W image is also shown 
for comparison. Note the different scales (HST/NICMOS: $40\arcsec \times 40 \arcsec$, 
Gemini/Hokupa'a: $20\arcsec \times 20 \arcsec$).}
\label{extihst}
\end{figure*}

Since the original HST/NICMOS field has twice the area of the Gemini
central Arches field, we have also calculated the extinction map for the entire
NICMOS field following the same procedure. The corresponding map is shown in 
Fig. \ref{extihst}.

The extinction measured with this procedure in the $K$-band lies in the range
$1.9 < A_K < 4.1$ mag, with an average value of 3.1 mag,
corresponding to $16 < A_V < 37$ mag, $<\!\!A_V\!\!> = 27.7$ mag.
Cotera et al. (\cite{Cotera2000}) derive a near-infrared extinction of $2.8 < A_K < 4.2$ mag,
with an average value of $<\!\!A_K\!\!>\ = 3.3$ mag for 15 lines of sight towards 
several Galactic Center regions, corresponding to an average visual extinction of
$<\!\!A_V\!\!>\ = 29.5$ mag (transformed using Rieke \& Lebofsky (\cite{RL85})). 
They obtain the highest extinction
towards a field close to the Arches cluster, $A_V = 37.5$ mag. 
This is very close to our highest extinction value.
The average value determined from individual dereddening here is the same
as the average extinction obtained by Figer et al. (\cite{FKM}), $<\!\!A_V\!\!>\ = 27.7$ mag.
Note that the typical random scatter $\sigma(A_K)$ from foreground dust density
fluctuations found in GC fields is linearly related to the average extinction
within a field. The relation determined by Frogel et al. (\cite{Frogel1999}) from giant 
branch stars in 22 pointings towards fields within 4\degr\ from the GC is given by 
$\sigma(A_K) = 0.056 (\pm 0.005)<\!\!A_K\!\!>\:+\ 0.043 (\pm 0.005)$. This yields an 
expected natural scatter from GC clouds of only 
$\sigma(A_K) = 0.22$ mag for $<\!\!A_K\!\!>\: = 3.1$ mag, much below the difference in 
reddening observed in the Arches field. Thus, the change in extinction cannot be
explained by the natural fluctuations of the dust distribution in the GC region.

Comparison of the cluster center main sequence population
with the main sequence colour of a theoretical 2 Myr isochrone
from the Geneva set of models (Lejeune \& Schaerer \cite{Lejeune2001}), later-on used for
the derivation of the mass function, yields an average extinction of $A_V = 24.1 \pm 0.8$ mag 
in the cluster center. 
This extinction value has been used to transform isochrone magnitudes
and colours into the cluster magnitude system. It has been suggested that the 
brightest and most massive stars in Arches are Wolf-Rayet stars of type WN7
(Cotera et al. \cite{Cotera1996}; Blum et al. \cite{Blum2001}).
Fundamental parameters of Wolf-Rayet stars are compiled in Crowther et al. 
(\cite{Crowther1995}).
For stars of subtype WN7 they find typical colours of $(H - K) \sim 0.2$ mag,
leading to an extinction of $A_V = 24.9 \pm 2.4$ mag with an observed 
$H - K$ colour of $\sim 1.77$ mag for the WN7 stars, which were 
identified by comparison with the Blum et al. (\cite{Blum2001}) narrow band photometry. 
This value is in very good agreement with the $A_V$ determined from the 
main sequence colour in the cluster center.

\subsection{Colour-magnitude diagrams}
\label{cmdsec}

\begin{figure}
\includegraphics[width=8cm,clip=]{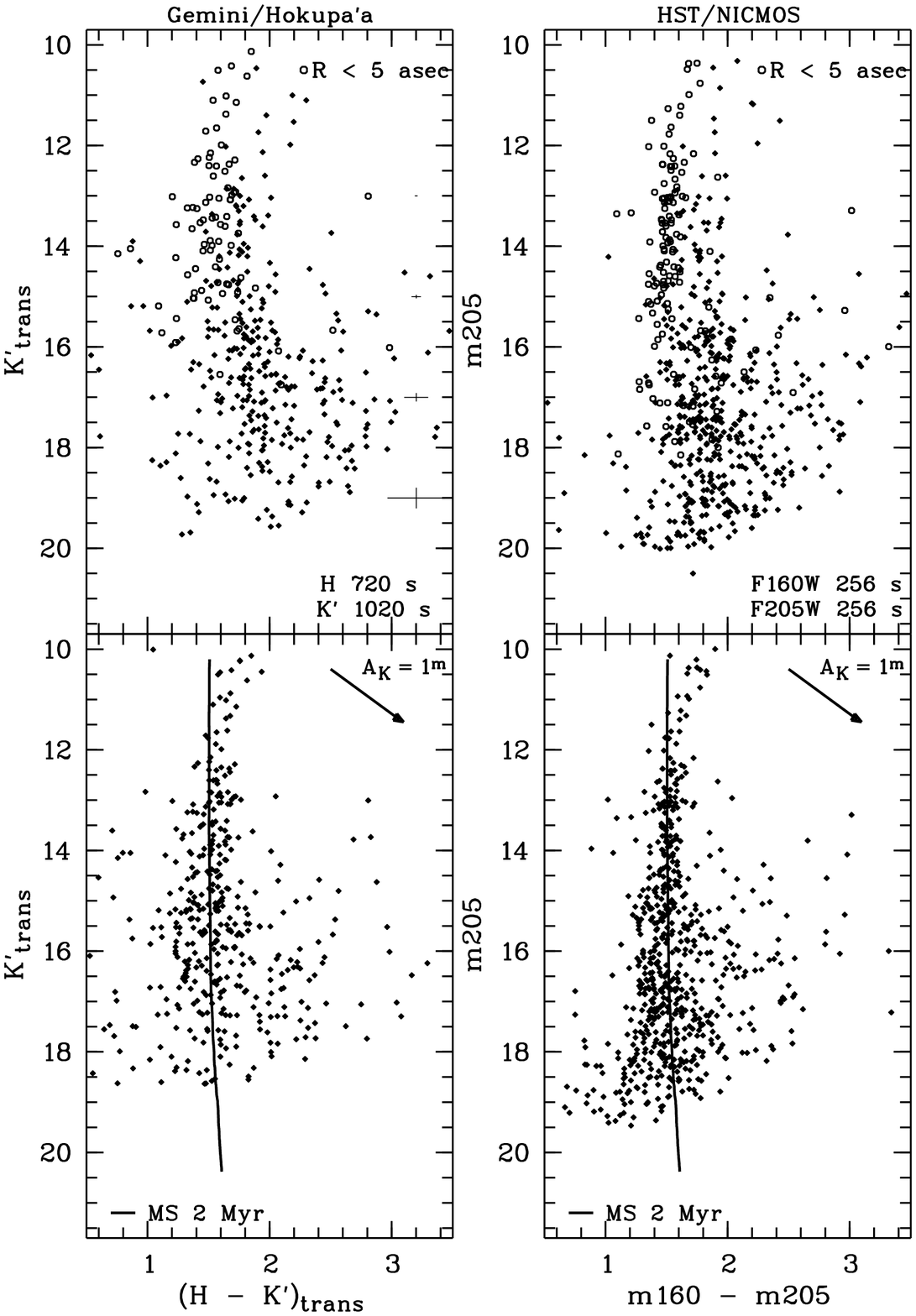}
\caption{Colour-magnitude diagrams. \newline
Left: Gemini/Hokupa'a, right: HST/NICMOS \newline
Lower panel: CMDs corrected for radial reddening gradient (Sect. \ref{trendsec})}
\label{cmd}
\end{figure}

The resulting colour-magnitude diagrams for Gemini and HST are presented in 
Fig. \ref{cmd} (upper panel). Two important differences are seen when inspecting the CMDs. 
First, the scatter in the main sequence is significantly larger in the ground-based 
photometry. While the HST/NICMOS CMD reveals a 
narrow main sequence in the cluster center (circles in Fig. \ref{cmd}), the same
stars display a much larger colour range in the Gemini CMD.
The poor Strehl ratio in the Gemini/Hokupa'a data as compared to 
the HST/NICMOS data (see Sect. \ref{strehlsec})
causes a high, non-uniform additional background due to uncompensated
seeing halos around bright stars, which decreases photometric accuracy.
In the dense regions of the cluster center, where crowding problems are most 
severe, the photometry is most affected. The number of faint, 
unresolved companion stars that merge into the high stellar background underneath
the bright cluster population is very high. As discussed in Sect. \ref{ressec}, 
the halos of the bright stars hinder the detection of faint objects 
despite the principally high spatial resolution seen in individual PSF kernels.
Operating at the diffraction limit,
NICMOS is not restricted by these effects, yielding a better effective resolution 
especially in the dense regions. A tighter main sequence and less
scatter is the consequence. 
In Fig. \ref{cmd}, the innermost 5\arcsec\ of the Arches
cluster are marked by open circles. 
It is clearly seen that most massive (bright) stars are located 
in the cluster center. 

A second effect observed is the much larger number of faint 
objects seen in the HST data (cf. Fig. \ref{magerr}). 
As the limiting magnitude and the measured spatial resolution of 
the images are similar in both datasets, this, too, 
has to be a consequence of the low Strehl ratio in the AO data. 

The lower panel of Fig. \ref{cmd} shows the ``dereddened'' CMDs, corrected
for the radial colour gradient found and the corresponding change 
in extinction, $\Delta A_K$ (Sect. \ref{trendsec}). The colours of
stars beyond $R > 5\arcsec$ have been adjusted to the colour of the cluster 
center. 
Comparison with the original CMDs shows that
most of the bright, seemingly reddened stars fall onto the same main sequence
after correcting for the colour trend. 
These stars are located at larger distances from the cluster
center and thus suffer from more reddening by residual dust. 
As will be discussed in the context of the mass function (Sect. \ref{mfsec}), 
these stars might have formed close to the cluster at a similar time 
as the cluster population. At the faint end of the CMD,
there are, however, a large number of objects that remain unusually red after
the correction has been applied. 
These objects may either be pre-main sequence stars or faint background sources.
Unfortunately, we are not able to disentangle these two possible contributions,
and will thus exclude objects significantly reddened relative to the main 
sequence when deriving the mass function.

\subsection{HST/NICMOS colour-colour diagram}
\label{ccdsec}

\begin{figure}
\includegraphics[width=5.5cm,angle=270,clip=]{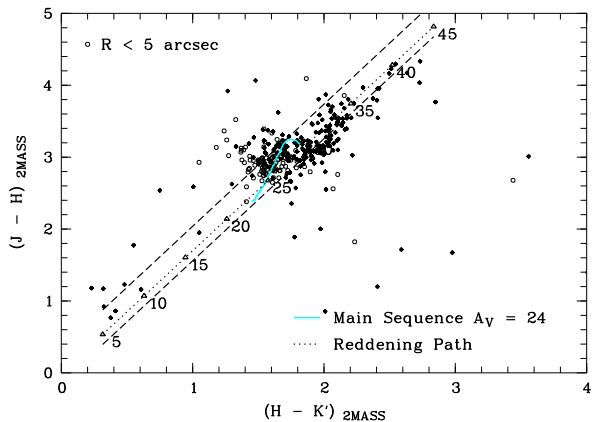}
\caption{Two-colour diagram from HST observations. The reddening path is shown 
as a straight line labeled with $A_V$ values, and the main sequence is indicated
by the thick grey line. The area between the dashed lines marks the region of
reddened main sequence stars.}
\label{ccd}
\end{figure}

For comparison with the reddening path and a main sequence in standard
colours, the NICMOS filters have been transformed into the 2MASS $JHK_s$ system.
In Fig. \ref{ccd}, we show the transformed HST/NICMOS colour-colour diagram
for the stars bright enough to be observed in all three filters.
The $A_V$ values are from the Rieke \& Lebofsky (\cite{RL85}) extinction 
law for standard $JHK$ photometry.
Though we are aware of the uncertainties inherent to the transformation
of severely reddened stars, the proximity of the reddening path to 
the data points supports the validity of the equations derived by BGB. 
Changing the transformation parameters slightly results in a large angle between the 
data points and the reddening path. 

A wide spread population of stars is clearly seen along the reddening path,
as expected from the colour trend discussed in Sect. \ref{trendsec}
(no correction for the varying extinction has been applied in this diagram).
Again, the stars with the lowest reddening within the cluster population 
are the bright stars in the Arches cluster center. 
Moving along the reddening line towards higher values of $A_V$
mainly means moving radially outwards from the cluster center.
As in the CMD, a correction for the observed colour gradient causes the bulk 
of the stars to fall onto the main sequence with a reddening of $A_V = 24$ mag,
corresponding to the cluster center. 

\section{Luminosity Functions and incompleteness effects}
\label{lfsec}

\subsection{Integrated luminosity function}
\label{lfallsec}

\begin{figure}
\includegraphics[width=5.5cm,angle=270,clip=]{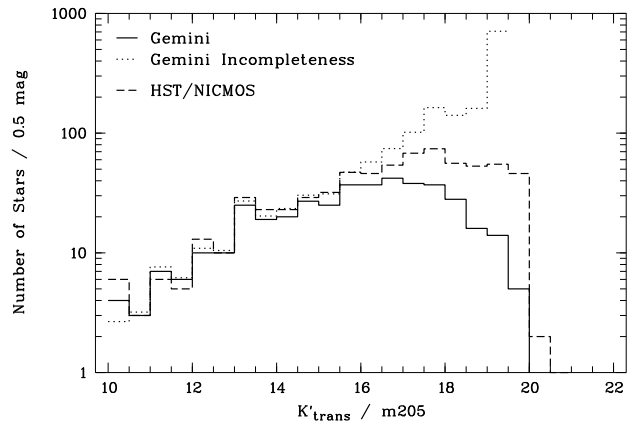}
\caption{Comparison of Gemini versus HST luminosity functions.}
\label{lf}
\end{figure}

In Fig. \ref{lf}, we compare the Gemini with the HST luminosity function for the
$K^\prime_{\rm trans}$ vs. $m205$ observations. 
For direct comparison of the observational efficiency, no colour 
cut has been applied, but the entire physically reasonable colour range
from approximately 0 to 4 mag in $H-K^\prime$, 
including reddened and foreground objects, has been included in the luminosity
function (LF). Therefore, these LFs are {\sl not} the ones from which the 
mass functions have 
been derived. The only selection criterion that has been applied
is a restriction of the photometric uncertainty in both magnitude 
($\sigma_{K^\prime} < 0.2$ mag)
{\sl and} colour ($\sigma_{H-K^\prime} < 0.28$ mag, 
corresponding to $\sigma_{K^\prime} < 0.2$ mag and $\sigma_H < 0.2$ mag). 
The uncertainty selection in colour allowed us to select only those 
objects that have been detected with high confidence in both $H$ and $K^\prime$
images in the Gemini data, and in the F160W and F205W filters in the NICMOS data,
respectively. The colour-uncertainty selection on the HST data simulates the 
matching of $H$ and $K^\prime$ detections used on the Gemini data 
for the selection of real objects. Thus, only objects that are detected
in both $H$ and $K^\prime$ have been included in the luminosity
function. This gives us 
some confidence that we are not picking up hot pixels or cosmic ray events.
The area covered with Gemini has been selected from the HST
photometry as displayed in Fig. \ref{images}.

As can be seen in Fig. \ref{lf}, many objects are missed
by Gemini in the fainter regime, though the actual limiting (i.e., cut-off)
magnitudes are the same in both datasets. 
This is due to the fact that 50 \% of the light is distributed into
a halo around each star. 
These halos prevent the detection of faint objects 
around bright sources, especially in the crowded regions. 
This effect is most obvious when examining the star-subtracted frames resulting 
from the DAOPHOT {\sl allstar} task. In these frames the cluster center is 
marked by a diffuse background, enhanced by $\sim 20$ counts in $K^\prime$ 
and $\sim 40$ counts in $H$ above the observational background of 
2 and 4 counts in the cluster vicinity, respectively.
In addition to the simple crowding problem due to the stellar density
affecting both datasets, 
the overlap of many stellar halos hinders the detection of faint 
 stars in the Gemini data. 
At larger radial distances from the cluster center, more and more faint stars
are detected both in the HST as well as in the Gemini data (Fig. \ref{lfkrad}).

The fact that the incompleteness corrected Gemini LFs follow closely the shape 
of the HST LFs supports the results of our incompleteness calculations, 
which will be used to determine the incompleteness in the mass function.

\subsection{Radial variation of the luminosity function}
\label{lfradsec}

\begin{figure}
\includegraphics[width=8cm,clip=]{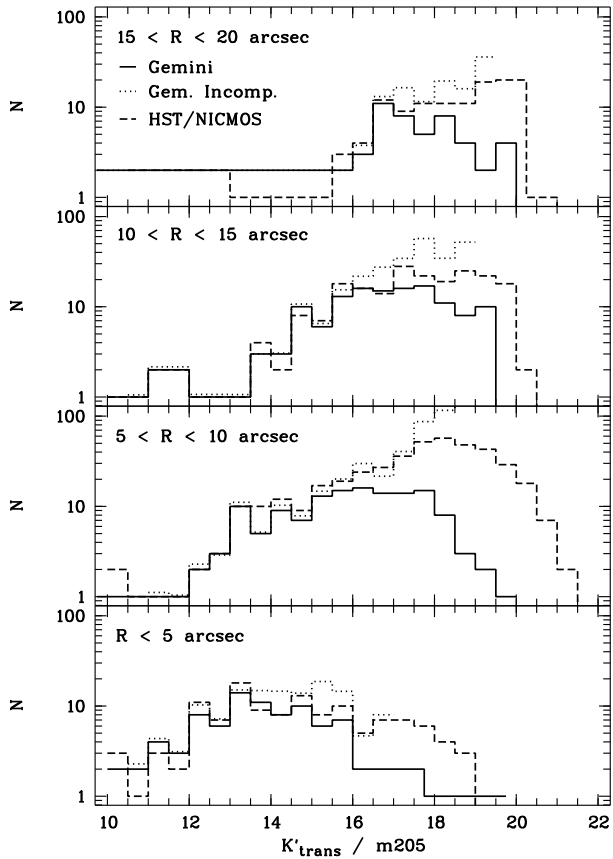}
\caption{Radial variation of the luminosity function.
The comparison of Gemini and HST LFs is shown together with the Gemini
incompleteness calculation. The dependence
of the magnitude limits on the distance to the cluster center is striking.}
\label{lfkrad}
\end{figure}

Radial luminosity functions were calculated in \mbox{$\Delta R = 5$ \arcsec} 
bins, using the same uncertainty selection as in Fig. \ref{lf} (Sect. \ref{lfallsec}).
The resulting radial LFs are shown in Fig. \ref{lfkrad}, along with the 
incompleteness determined for each radial bin.
In the cluster center (lowest panel), the very good match of the Gemini and
HST LFs for $K^\prime_{\rm trans} < 16$ mag reveals the comparable spatial 
resolution obtained in both data sets. Despite the strong crowding seen
already in these bright stars, the Gemini AO data resolve the sources 
in the cluster center nicely. When we move on to fainter magnitudes, however,
we are limited by the halos of these bright stars, as discussed above.
The clear decrease below $K^\prime_{\rm trans} = 16$ mag marks
the point where stars are lost due to the enhanced background. When we move
radially outwards, the limiting magnitude above which faint stars are lost
shifts towards fainter magnitudes. The tendency to loose the faint tail
of the magnitude distribution nevertheless remains clearly seen, though
it becomes much less pronounced for $R > 10\arcsec$, where the Gemini and 
HST LFs resemble each other. For $R > 15\arcsec$ (upper panel) we are limited by
small number statistics due to the small area in this radial bin.
As it is hard to observe a well-defined LF at these radii, we will add the
two upper bins when we create the radially dependent mass functions in 
Sect. \ref{mfradsec}.

The magnitude-dependent distribution of stars within the cluster is evident in 
these LFs. While bright stars are predominantly found in the cluster center, 
their number density strongly decreases with increasing radius.
When we analyse the Gemini LFs more quantitatively, 
we find 25 (50) stars with $K^\prime_{\rm trans} < 13 (14)$ mag within $R < 5\arcsec$, but
only 8 (23) such stars with $5 < R < 10\arcsec$, and beyond 10\arcsec,
we observe only 7 (11) such stars. The numbers for HST are comparable in the bright 
magnitude bins. On the other hand, the number of faint stars with 
$K^\prime_{\rm trans} > 18 (19)$ mag increases from 1 (0) to 14 (3) to 48 (16). 
As we see significantly more faint stars in the HST data, the corresponding 
numbers are higher, i.e. the number of stars with $m205 > 19$ mag is 0 in the 
innermost bin, 24 in the intermediate bin, and 81 in the outermost bin.
Despite the fact that the area on the Gemini frame increases by about a factor 
of 3 between the inner and intermediate bin, the number of bright stars is strongly 
diminished beyond a few arcseconds, while the number of the faint stars increases
by much more than the change in area can account for. Although for the fainter stars
the effects of crowding and a real increase in the fainter population of the 
cluster cannot be disentangled, the decrease in the number of bright stars is 
a clear indication of mass segregation within the Arches cluster.

\section{Mass Function}
\label{mfsec}

The mass function (MF) may be defined as the number of stars observed in a certain 
mass bin. The mass function in stellar populations is most frequently
fitted by a power law, whose slope depends on the mass range analysed
(e.g, Kroupa \cite{Kroupa2001}).
In the logarithmic representation, the mass function is defined as
\begin{center}
\vspace{-3ex}
$$ {\rm d} (\log N)/{\rm d} (\log M) \sim M^\Gamma $$
\end{center}
where $M$ is usually given in solar masses, and $\Gamma$ is the slope of the 
mass function. The exact shape of the mass function and, in particular,
its slope, has been subject of intense discussion (see, e.g., Massey \cite{Massey1995a},
\cite{Massey1995b}; Scalo \cite{Scalo1986}; and Scalo \cite{Scalo1998} for a review).
Massey et al. (\cite{Massey1995a}, \cite{Massey1995b}) report slopes
between $-0.7$ and $-1.7$ with a weighted mean of $-1.1 \pm 0.1$ for $M > 7\ M_\odot$
in several starforming regions in the Milky Way, and $-1.1 < \Gamma < -1.6$
with a mean of $-1.3 \pm 0.1$ for $M > 25\ M_\odot$ in the LMC. 
From these studies in young clusters and associations, the mass function
has been suspected to be rather universal, following a Salpeter (1955) power law
with a slope of $\Gamma = -1.35$ for stars with masses 
$M \ge 1\,M_\odot$ (Kroupa \cite{Kroupa2001}). 
For the two starburst clusters studied in the Milky Way, Arches (FKM) and NGC\,3603
(Eisenhauer \cite{Eisenhauer1998}), a flat slope of $\sim -0.7$ is observed
for masses $M > 10 M_\odot$ in Arches, and $M > 1 M_\odot$ in NGC\,3603.
These values have to be kept in mind as a comparison for the results presented
in the following sections. The mass regime which will be discussed here 
spans a mass range of $6\!<\!M\!<\!65\ M_\odot$ for Arches cluster stars.

\subsection{Integrated mass function}

\begin{figure}
\includegraphics[width=6cm,angle=270,clip=]{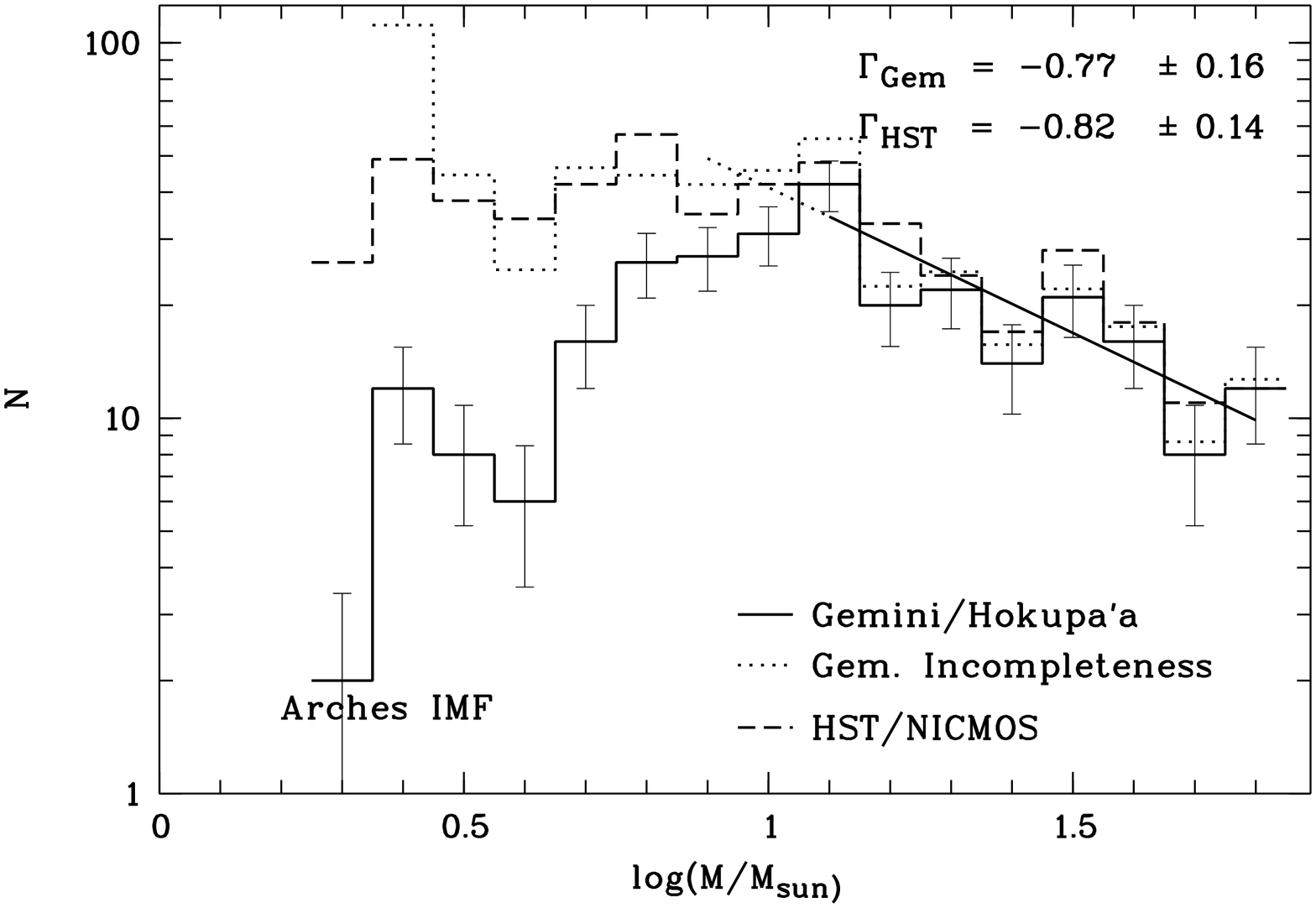}\\
\\
\includegraphics[width=6cm,angle=270]{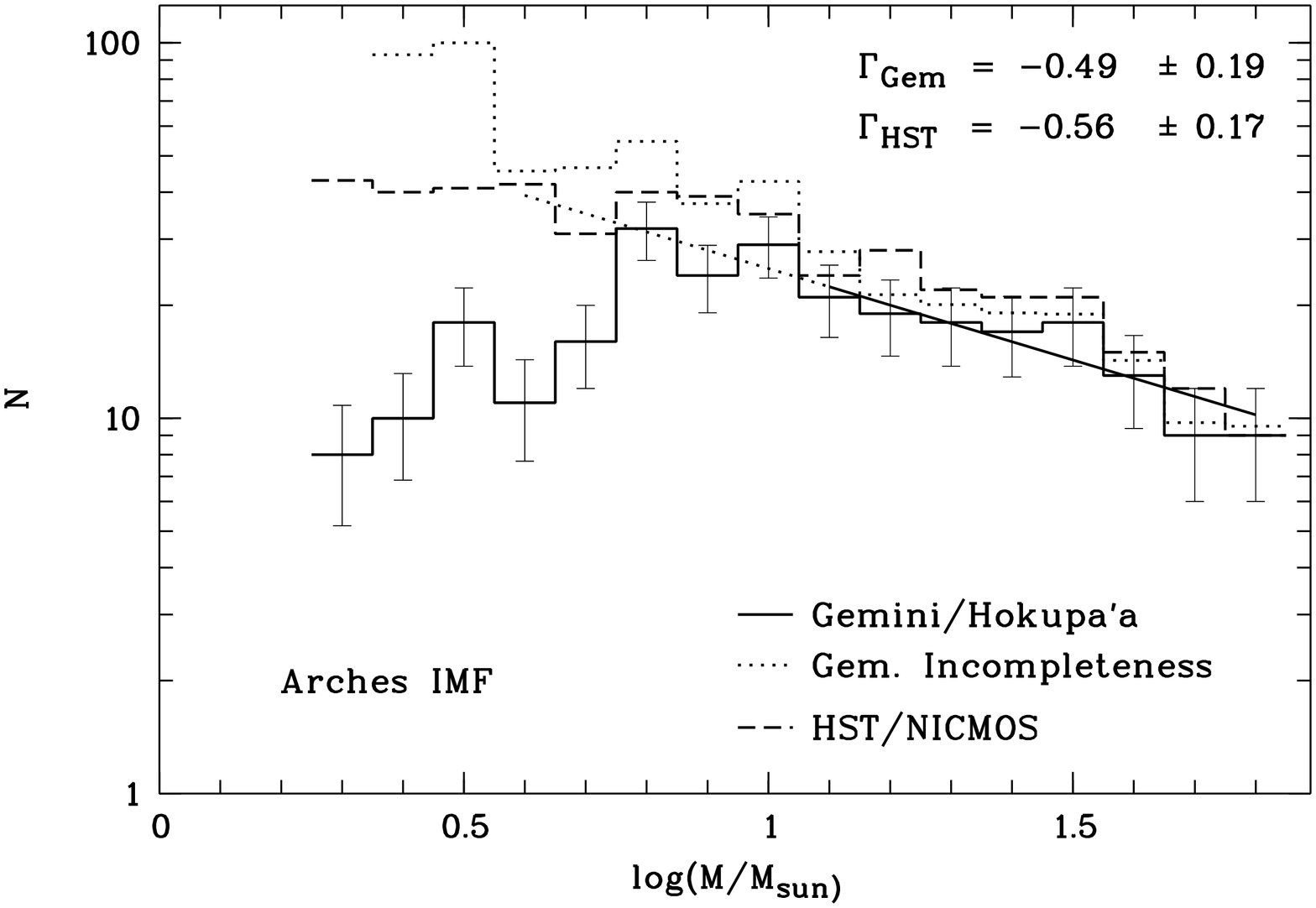}
\caption{Arches mass function derived from the Gemini/Hokupa'a colour-magnitude
diagram shown in Fig. \ref{cmd}. A 2 Myr main sequence
isochrone from the Geneva set of models (Lejeune \& Schaerer \cite{Lejeune2001})
was used to transform magnitudes into stellar masses. 
The mass function has been derived for bins of $\delta \log (M/M_\odot) = 0.1$
with a lowest mass bin $\log (M/M_\odot) = 0.20$. \newline
Upper panel: with $A_K$ correction \newline
Lower panel: without $A_K$ correction \newline}
\label{mf}
\end{figure}

The present-day mass function (Fig. \ref{mf}) of the Arches cluster 
has been derived from the colour-magnitude diagram
by transforming stellar luminosities into masses via a 2 Myr isochrone
from the Geneva basic set of stellar evolution models 
(Lejeune \& Schaerer \cite{Lejeune2001})
using the method described in Grebel \& Chu (\cite{GC2000}).
Enhanced mass loss models were also used, but did not alter the resultant 
mass function. Stellar evolution (mass loss, giant evolution)
is not important on timescales of the Arches age of $\sim 2$ Myr
for stars with initial masses of $M < 50\,M_\odot$ ($\log (M/M_\odot) < 1.7$,
i.e. stellar evolution affects the two upper mass bins of the MF at most).
No attempt has been made to reconstruct the initial mass function (IMF)
from the present-day MF for $M > 50\,M_\odot$. 
A distance modulus of 14.5 mag and an extinction of $A_V = 24.1$ mag
have been applied. 

The slopes of all mass functions discussed have been derived by performing a
weighted least-squares fit to the number of stars per mass bin. 
The size of the mass bins was chosen to be $\delta\log (M/M_\odot) = 0.1$ as the best
compromise between mass function resolution and statistical relevance.
This bin size is significantly larger than the photometric uncertainty in 
the considered mass and thus magnitude ranges. Only mass bins 
with a completeness factor of $\ge 75$\% have been included in the fit.

Note that we have not attempted to subtract the field star contribution.
As can be seen in Fig. \ref{images}, the Gemini field is mostly restricted to 
the densest cluster region. When comparing to an arbitrary part of the GC field,
we do not expect to observe the same distribution of stars as in the Arches field,
as the faint, reddened background sources are negligible due to the high density of 
bright sources in the cluster area.

In addition, the stellar density in the GC is strongly variable,
imposing additional uncertainties on the field contribution. 
Neither the Gemini nor the HST field covers enough area to estimate
the field star population in the immediate vicinity of the cluster.
A main sequence colour cut ($1.15 < H-K < 1.90$ mag) has been applied to the colour-corrected CMDs 
to select Arches members, excluding blue foreground and red background sources. 

To allow for a direct comparison with the results obtained in FKM, 
we have used isochrones calculated for a metallicity of $Z = 0.04$ for all
MF derivations. The derived MFs are displayed in Fig. \ref{mf}.
The overall mass function derived from the Gemini data displays the same
slope as derived from NICMOS within the uncertainties, namely $\Gamma_{\rm Gemini} = -0.77 \pm 0.16$
and $\Gamma_{\rm HST} = -0.82 \pm 0.14$ fitted for $10\!<\!M\!<\!65\ M_\odot$ (Fig. \ref{mf}),
which may be extrapolated down to $6\ M_\odot$ when taking into account the incompleteness
correction. The present-day upper mass of $65\,M_\odot$ corresponds to an {\sl initial}
mass of about $100\,M_\odot$ according to the Geneva models.
FKM derive an overall slope of $\Gamma = -0.7 \pm 0.1$ in the inital mass range
$6\!<\!M\!<\!120\ M_\odot$, in good agreement with our results.
The remaining difference in the maximum mass is due to the different extinction
and the extinction corrections applied, which represent the largest 
uncertainties in the mass function derivation.
As in particular the correction of the $K$ magnitude for the varying extinction
is uncertain due to the unknown extinction law of the NICMOS filters, we have
also derived the mass function for uncorrected $K$ magnitudes, with only the 
colour correction applied, which is independent of the extinction law assumed.
In this case, the MF appears flatter with a slope of $\Gamma \sim -0.5 \pm 0.2$ 
(Fig. \ref{mf}, lower panel).
The discrepancy in the derived slopes clearly shows that the effects of 
differential extinction are not negligible, especially when deriving mass 
functions for very young regions, where the reddening varies significantly.

Furthermore, we have checked the effect of the binning on the MF by shifting
the starting point of each bin by one tenth of the bin-width, 
$\delta\log (M/M_\odot) = 0.01$. The resultant slopes range from 
$-0.69 \pm 0.13 < \Gamma_{\rm Gemini} < -0.90 \pm 0.15$ and
$-0.79\pm 0.12 < \Gamma_{\rm HST} < -0.98 \pm 0.13$. The average slopes for Gemini and
HST, $\Gamma_{\rm Gemini} = -0.77 \pm 0.15$ and $\Gamma_{\rm HST} = -0.86 \pm 0.13$,
respectively, agree well within the errors.
The slightly flatter slope observed in the Gemini data may reflect the more severe
incompleteness due to crowding.
Although all slopes are consistent within the errors,
the range in slopes derived by scanning the bin-step shows that 
statistical effects due to the binning may not be entirely neglected in the MF derivation.

The metallicity within the immediate Galactic Center region has been a matter
of discussion during the past decade. Several authors report supersolar
metallicities derived from CO index strength and TiO bands in bulge stars 
(Frogel \& Whitford \cite{Frogel1987}; Rich \cite{Rich1988}; Terndrup et al. 
\cite{Terndrup1990}, \cite{Terndrup1991}). 
Carr et al. (\cite{Carr2000}) measure [Fe/H] $= -0.02 \pm 0.13$ dex for the 7 Myr old
supergiant IRS 7, and Ramirez et al. (\cite{Ramirez2000}) derive an average of
[Fe/H] $= +0.12 \pm 0.22$ dex for 10 young to intermediate age supergiants, both 
very close to the solar value. 

Using a 2 Myr isochrone with solar metallicity Z=0.02, 
the average slope of all bin steps is $\Gamma_{\rm Gemini} = -0.84 \pm 0.13$ and
$\Gamma_{\rm HST} = -0.91 \pm 0.12$.
The mass function is thus not significantly altered when using solar
instead of enhanced GC metallicity. We note, however, that a lower metallicity
(i.e., in this case solar) steepens the MF slightly, thus working into the same
direction as the incompleteness correction. 

FKM report a flat portion of the
MF in the range $15\!<\!M\!<\!50\ M_\odot$, which is not seen in the Gemini
MF. This plateau can, however, be recovered, when we create a MF from 
$K$-band magnitudes uncorrected for differential extinction, 
and use a lowest mass of $\log (M_{\rm low}/M_\odot)= 0.25$.
For $\log (M_{\rm low}/M_\odot) = 0.20$ the plateau is seen neither with 
nor without extinction correction.
This, again, shows the dependence of the shape of the MF on the extinction corrections
applied, as well as on the chosen binning. 

From the considerations above, 
we conclude that the overall mass function of the Arches cluster has 
a slope of $\Gamma = -0.8$ to $-0.9$ in the range $6\!<\!M\!<\!65\ M_\odot$.
Although the uncertainty of missing lower mass stars in the
immediate cluster center remains, the incompleteness correction strongly supports 
the derived shape of the MF. If the flat slope would be solely due to a low recovery
rate of low-mass stars in the cluster center, this should be visible in a much 
steeper rise of the incompleteness corrected MF in contrast to the observed MF.
We thus conclude that the slope of the MF observed in Arches is flatter than the 
Salpeter slope of $\Gamma = -1.35$, assumed to be a standard mass distribution in 
young star clusters. Such a flat mass function is a strong indication of the 
efficient production of high-mass stars in the Arches cluster and the GC environment. 

\subsection{Effects of the chosen isochrone, bin size, and metallicity}

Blum et al. (\cite{Blum2001}) estimate a cluster age of $2-4.5$ Myr for Arches
assuming that the observed high-mass stars are of type WN7. 
If we compare the Geneva basic grid of isochrones with fundamental
parameters obtained for WN7 stars (Crowther et al. \cite{Crowther1995}), a reasonable
upper age limit for this set of isochrones is $\sim 3.5$ Myr.
Crowther compares the parameters derived for WN stars with evolutionary
models at solar metallicity from Schaller et al. (\cite{Schaller1992}) and with the 
mass-luminosity relation for O supergiants from Howarth \& Prinja (\cite{Howarth1989}), 
yielding a mass range of $20\,-\,55\,M_\odot$, but with high uncertainties
at the low-mass end. The more reliable mass estimates for the colour 
and magnitude range observed for WN stars in Arches are restricted to 
$35\!<\!M\!<\!55\ M_\odot$. From spectroscopic binaries, the masses
of two WN7 stars are determined to be $\sim 30\ M_\odot$ and $> 48\ M_\odot$
(Smith \& Maeder \cite{Smith1989}).
The theoretical lower limit to form Wolf-Rayet
stars is $25\ M_\odot$ for the Geneva models 
(Schaerer et al. \cite{Schaerer1993}). 

In the Geneva basic grid of models with $Z = 0.04$,
the 3.5 Myr isochrone is limited by a turnoff mass of $32\,M_\odot$.
We have thus calculated mass functions for isochrones with ages 2.5, 3.2, and
3.5 Myr in addition to the 2 Myr case discussed above. 
Though the derived slopes scatter widely, irrespective of the isochrone 
used, the slope of the MF tends to be even {\sl flatter} for any of the older
population models. 
We thus conclude that, regardless of the choice of model and 
parameters, the Arches mass function displays a flat slope.

\subsection{Radial variation in the mass function}
\label{mfradsec}

\begin{figure}
\includegraphics[width=8cm,clip=]{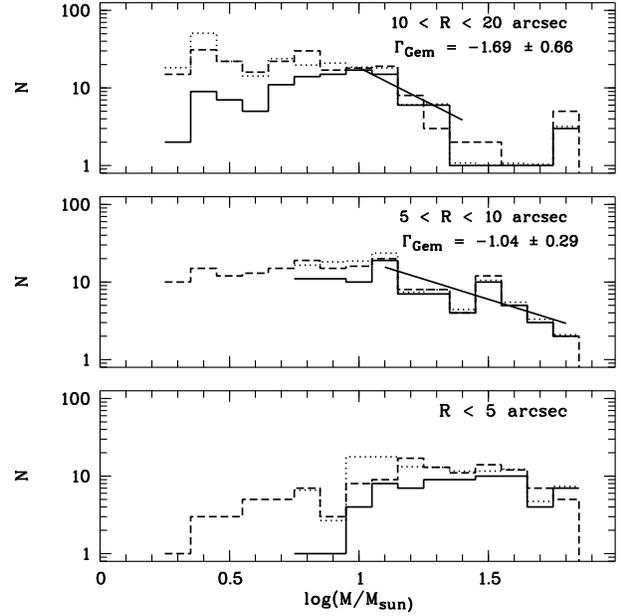}
\caption{Radial change of the mass function as observed in the Gemini/Hokupa'a 
data. A very flat mass function is seen in the inner cluster regions, where
predominantly massive stars are found. The slope of the mass function 
increases towards the Salpeter value ($\Gamma = -1.35$) already at a radial
distance from the cluster center of $>$ 10\arcsec\ (0.4 pc at a 
distance of 8 kpc).}
\label{mfrad}
\end{figure}

The radial variation (Fig. \ref{mfrad}) of the mass function is 
particularly interesting with respect to YC evolution. 
We have analysed the stellar population in Arches within
three different radial bins, $0\arcsec < R < 5\arcsec$, 
$5\arcsec < R < 10\arcsec$ and $10\arcsec < R < 20\arcsec$.
The resulting mass functions for the Gemini and HST datasets, 
along with the radius dependent incompleteness correction for the Gemini MFs, 
are displayed in Fig. \ref{mfrad}.

We confirm the flat mass function observed by FKM
in the innermost regions of the cluster, which steepens
rapidly beyond the innermost few arcseconds. 
Most of the bright, massive stars are
found in the dense cluster center, where the mass function slope is very flat.
FKM derive a slope of $\Gamma = -0.1 \pm 0.2$ from the HST data in this region,
which is consistent with Fig. \ref{mfrad}. It is obvious from the lowest panel 
in Fig. \ref{mfrad} that we are crowding limited in the innermost region.
We have thus not tried to fit a slope for $R < 5\arcsec$.

In the next bin, $5\arcsec < R < 10\arcsec$, the mass function obtained from 
the weighted least-squares fit has already steepened to a slope of $\Gamma = -1.0 \pm 0.3$.
Again, the MF in this radial bin remains significantly flatter ($\Gamma = -0.5 \pm 0.4$)
when no $A_K$-correction is applied.
Beyond 10\arcsec\ (0.4 pc, upper panel), a power law can only be defined in the range
$10\!<\!M\!<\!30\ M_\odot$ ($1.0 < \log (M/M_\odot) < 1.5$), where a slope of 
$\Gamma = -1.69 \pm 0.66$ is found, consistent with a Salpeter ($\Gamma = -1.35$) law. 
The large error obviously reflects the small
number of mass bins used in the fit. Nevertheless, Fig. \ref{mfrad} clearly 
reveals the steepening of the MF very soon beyond the cluster center.

\begin{figure}
\includegraphics[width=7cm,height=8cm,angle=270,clip=]{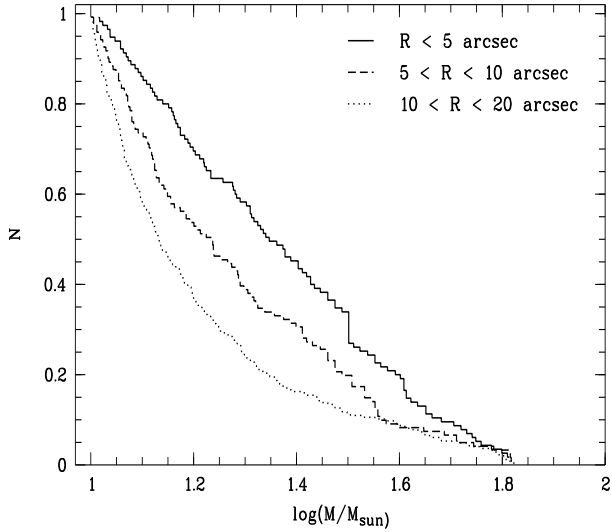}
\caption{Cumulative functions of the stars in each radial bin obtained from 
HST/NICMOS data. Each radial bin corresponds to one radial mass function in 
Fig. \ref{mfrad}. The cumulative distributions have been normalised at $10\,M_\odot$,
the limit where all radii are more than 75 \% complete.}
\label{cum}
\end{figure}

For a more quantitative confirmation of the mass segregation
present in the Arches cluster, we created cumulative functions for the 
mass distributions in the three radial bins (Fig. \ref{cum}).
We have applied a Kolmogorov-Smirnov test to quantify the observed differences
of these functions. 
When the central, intermediate, and outer radial bin are compared pairwise,
we obtain in each case a confidence level of more than 99 \% that the 
mass distributions do not originate from the same distribution.

Thus, the inner regions of the cluster are indeed skewed
towards higher masses either by sinking of the high mass stars towards the cluster 
center due to dynamical processes or by primordial mass segregation, or both. 
The same effect is observed in the 
similarly young cluster NGC\,3603 (Grebel et al. 2002, in prep.).

\subsection{Formation locus of massive stars}
\label{formsec}

We find several (5)
high mass stars in the cluster vicinity. These stars fall onto the Arches main 
sequence after applying the reddening correction (Sect. \ref{trendsec}).
When the entire HST field is separated into two equalsize areas, 
the first area being a circle of radius 16\arcsec\ around 
the cluster center, and the second the surrounding field, 
no additional comparably high-mass stars are found with Arches main sequence colours
except for the two bright stars found at the edges of the Gemini field. 
In the dynamical models of
Bonnell \& Davies (\cite{Bonnell1998b}) there is a low, but non-zero probability that
a massive star originating outside the cluster's half-mass radius might
remain in the cluster vicinity. High-mass stars formed near the center
show, however, a tendency to migrate closer to the cluster center. 
The disruptive GC potential might enhance the ejection process of low-mass stars, 
but according to equipartition, it is unlikely that the most massive stars gain 
energy due to dynamical interaction with lower mass objects. 
We have to bear in mind, however, that interactions between massive stars in the 
dense cluster core could cause the ejection of high-mass stars. 

N-body simulations performed
by Portegies Zwart et al. (\cite{PZ1999}) show that the inclusion
of dynamical mass segregation in cluster evolution models enhances the 
collision rate by about 
a factor of 10 as compared to theoretical cross section considerations.
For a cluster with 12000 stars initially distributed according to a Scalo 
(\cite{Scalo1986}) 
mass function, a relaxation time of 10 Myr, and a central density and half-mass radius
comparable to the values found in Arches, about 15 merging collisions occur
within the first 10 Myr ($1\,t_{\rm relax}$) of the simulated cluster. Shortly after
the start of the simulations, frequent binary and multiple systems form from 
dynamical interactions leading to the ejection of several contributing massive
stars. The flat MF in the Arches center as compared to a Scalo MF, 
containing a larger fraction of massive stars to interact, may even
increase the collision rate.

Though it is likely that the high-mass stars seen in the immediate vicinity 
of the Arches cluster have formed from the same molecular cloud at the same time as the 
cluster, a final conclusion on the possible ejection of these stars from the 
cluster core due to dynamical processes can only be drawn when velocities 
for these cluster member candidates are available.

\section{Comparison with cluster formation models}

In this section, we will first use the simple analytical approach 
to analyse internal cluster dynamics as summarised in Binney \& Tremaine (1987), 
and will then compare the derived dynamical timescales to N-body simulations
where the tidal force of the Galactic Center potential is considered.

From the transformation of stellar magnitudes into masses, a rough estimate
of the timescales relevant for dynamical cluster evolution can be made.
For larger area coverage, we have used the HST/NICMOS data in the calculation
below. The timescales characterising cluster evolution are the cluster's crossing
time, $t_{\rm cross}$, which is simply given by some characteristic radius divided
by the average velocity, i.e., the mean velocity dispersion of the cluster, 
$r_{\rm c}/\langle \sigma \rangle$, and the relaxation time, $t_{\rm rh}$.
The median relaxation time is
the time after which gravitational encounters of stars have caused the system 
to equilibrate to a state independent of the original stellar orbits 
(Binney \& Tremaine \cite{BT87}, hereafter BT87) 
\begin{displaymath}
t_{\rm rh} = \frac{6.63 \cdot 10^8 {\rm yr}}{\ln(0.4N)}\left(\frac{M}{10^5\,M_\odot}\right)^{1/2}\left(\frac{1\,M_\odot}{m_\ast}\right)\left(\frac{r_{\rm c}}{1\,{\rm pc}}\right)^{3/2}, 
\end{displaymath}
where $M$ is the total mass within some characteristic radius $r_{\rm c}$, 
$m_\ast$ is a characteristic stellar mass, here defined as the median of
the observed mass distribution, and $N$ is the number of stars in the cluster. 
$t_{\rm rh}$ may also be expressed in terms of the
crossing time, $t_{\rm rh} \sim 0.1\,N/\ln N \cdot t_{\rm cross}$. 

To evaluate the above formula, we need to know the total mass of the system, $M$,
a characteristic radius, $r_{\rm c}$, the number of stars, $N$,
and the characteristic stellar mass, $m_\ast$.
In principle, we are able to derive most of these quantities from 
isochrone fitting, by assigning each star a mass corresponding to its
$K$-band luminosity, and analysing the resulting spatial distribution of masses.
Naturally, these individual masses cannot be accurate for each star, as they 
depend strongly on the choice of the isochrone and inherit the
uncertainties of the photometry. The integrated properties are, however,
not very sensitive to the age of the model isochrone used, within the reasonable
age range of Arches, $\sim 2-3$ Myr (results will be given below). 
We have attempted to create a density profile from the HST/NICMOS data,
but as the profile appears to be very distorted, we have chosen to use
the half-mass radius, $r_{\rm hm}$, as a characteristic scale. 
$r_{\rm hm}$ has been obtained from the observed 
stellar population of the cluster under the assumption that the mass in the 
cluster center, relevant for the spatial scale on which gravitational 
interactions
are important, is dominated by the detected high-mass stellar population 
in the cluster center.
The relaxation time derived from $r_{\rm hm}$ is usually referred to as the
``half-mass relaxation time'' (cf. Portegies Zwart et al. \cite{PZ2002}).

The mass distribution within the cluster as derived from the isochrone allows
us to estimate $r_{\rm hm}$, $M$, and $m_\ast$. Taking all stars within 
a certain radius with Arches main sequence colours as cluster members,
we can also estimate $N$. 

We obtain a half-mass radius of
$r_{\rm hm} = 10\arcsec\,\hat{=}\,0.4$ pc, irrespective of the age of the isochrone
used (2 Myr or 3.2 Myr) to transform $K$-band magnitudes into masses.
The total mass measured within this radius is $M(r_{\rm hm}) = 6300\,M_{\odot}$
($5400\,M_{\odot}$ for the 3.2 Myr isochrone).
By extra\-polation of the mass function down to $0.1\ {\rm M_{\odot}}$, FKM
derived a total mass of $12000\,M_{\odot}$ within 9\arcsec\ radius.
To estimate the total mass in the cluster center, they used the relative fraction
of stars with $M > 40\,M_{\odot}$ in two annuli separated at $R = 3\arcsec$
as a scaling factor. This procedure ignores the segregation of high-mass stars.
As the cluster shows strong evidence for mass segregation (see Sect. \ref{mfradsec}),
this may overestimate the total cluster mass (within $R < 9\arcsec$).
On the other hand, our mass estimate, ignoring incompleteness and the lower
mass tail, is a lower limit to the total mass within $R < 10\arcsec$.
We thus conclude that for an order-of-magnitude estimation,
a total mass of $10^4\,M_{\odot}$  within the half-mass radius 
is a good approximation. This yields an average mass density of 
$\rho(R<r_{\rm hm}) \sim 4\cdot 10^4\,M_\odot\,{\rm pc}^{-3}$. 

The calculation of the relaxation time requires an estimate of the characteristic
stellar mass and the number of stars. The median mass within $R < 10\arcsec$ 
is $7.3\ {\rm M_{\odot}}$ in our observed mass range of $1.4 < M < 65\,M_\odot$ 
in the HST sample, and the number of stars actually observed is 486.
We choose to use the median mass here as a more realistic mass estimate
than the mean for individual stars participating in interaction processes, as
we are aware of the fact that the mean is highly biased by the high fraction 
of high-mass stars in the cluster center. Nevertheless, we have to keep in mind
that this median value is still unrealistically high as we miss the low-mass
tail of the mass distribution, which also has the highest number of stars.

As the low-mass tail of the distribution is highly incomplete, we estimate $N$ 
to be at least $10000\,M_\odot/7.3\,M_\odot \sim 1400$.
This results in $t_{\rm rh} = 1.2$ Myr. Decreasing the characteristic mass and
increasing the number of stars lengthens the relaxation time.
For instance, lowering $m_\ast$ to $1\,M_\odot$, where we take into account
the fact that the cluster center is mass segregated as opposed to a standard 
Salpeter mass function and thus the characteristic mass of a star should be higher, and 
raising N to 10000 stars, would result in $t_{\rm rh} = 6.2$ Myr.
If a significant fraction of low-mass stars would have been formed and survived 
in the cluster center, the half-mass relaxation time would increase even more.
The present-day relaxation time is thus at least on the order of or 
longer than the age of the Arches cluster.

A relaxation time longer than the lifetime of the cluster would imply that 
not yet sufficient time has passed for dynamical mass segregation
to take place and would thus indicate that primordial mass segregation
was present at the time of cluster formation.
However, we have so far ignored the Galactic Center tidal forces 
accelerating the dynamical evolution of the cluster.
Kim et al. (\cite{Kim2000}) performed N-body simulations using the 
observed parameters of the Arches cluster to trace the dynamical
evolution of the cluster and constrain initial conditions.
They use a total mass of $2 \cdot 10^4 M_\odot$, tidal radius of 
$\sim 1$ pc and a single power-law IMF with slopes of
$\Gamma = -0.5,-0.75,-1.0$ and $-1.35$.
The number of stars ranges between 2600 for a lower mass cutoff of
$m_{\rm low} = 0.1\,M_\odot$ and 12700 for $m_{\rm low} = 1\,M_\odot$. From comparison 
with the HST/NICMOS data of FKM their models yield a power law with 
$\Gamma = -0.75$ as the most probable initial mass distribution.
Although this slope is very close to the observed present-day MF,
they also note that mass segregation takes place on timescales as short as
1 Myr, such that the cluster looses all memory of the initial conditions 
shortly after formation. Portegies Zwart et al. (2002) use a Scalo
(1986) MF to model the Arches cluster, 
and derive a total cluster mass of $4 \cdot 10^4 M_\odot$,
and an initial relaxation time of 20-40 Myr. When comparing with the 
same set of HST/NICMOS observations, they conclude that a standard IMF can evolve 
into the current MF due to dynamical segregation, and that the IMF did 
not need to be overpopulated in high-mass stars.
Their computations indicate that the relaxation time
is strongly variable. During the dynamical evolution, $t_{\rm rh}$ increases strongly
after core collapse, which occurs within $\sim 2$ Myr, due to cluster
re-expansion, and only starts to decrease after 10 Myr or later.
The observed relaxation time does thus probably not 
trace the cluster's initial conditions, but reflects the current dynamical
state in the evolution of Arches. 
In particular, a present-day relaxation time larger than the cluster's age
does not necessarily imply that the cluster is not dynamically relaxed.

Unfortunately, this means that we are not able to distinguish between 
primordial and dynamical mass segregation from the estimated relaxation time.
For the Orion Nuclear Cluster (ONC) and its core, the Trapezium, which has been
studied in greater detail (e.g., Hillenbrand 1997, Hillenbrand \& Hartmann 1998), 
Bonnell \& Davies (\cite{Bonnell1998b}) derive dynamical timescales
too long to explain the segregation observed in high-mass stars within the 
Trapezium by dynamical evolution, concluding that a significant
amount of primordial segregation must have been present. In the case of the Arches
cluster, the external gravitational field acts towards a fast disruption,
thereby impeding the equilibration process, such that dynamical segregation
may well be under way. 

As the different model calculations do not agree with respect to the IMF
required to create the observed present-day mass distribution, a final conclusion
on whether or not the Arches {\sl initial} mass function had to be enriched in 
massive stars, thus supporting high-mass star formation models, cannot be drawn.

The evaporation time, setting the scale for dynamical dissolution 
by internal processes, can be estimated as $t_{\rm evap} \approx 136\,t_{\rm relax} \sim 177$ Myr 
(857 Myr for $m_\ast = 1\,M_\odot$, BT87).
Kim et al. (\cite{Kim1999}) have shown that the time required to disrupt an Arches-like 
cluster within the GC potential is only 10 Myr, much shorter than evaporation 
by internal processes will ever be relevant. The models by Portegies Zwart et al.
({\cite{PZ2002}) suggest somewhat longer evaporation times in the range 
$30 < t_{\rm evap} < 50$ Myr for an Arches-like cluster.
The external potential is thus the dominating
factor in the dynamical evolution of Arches. Note that, however, even a cluster 
as dense as Arches would not survive for more than 1 Gyr independent of its
locus of formation. The relaxation and evaporation times found for Arches are 
very similar to our results for the young, compact cluster in \object{NGC\,3603} 
(Grebel et al. 2002, in prep.). 
However, as NGC\,3603 is not torn apart by additional external tidal forces, 
it may survive much longer than Arches.
For comparison, massive young clusters in the Magellanic Clouds
have relaxation times of $\sim 10^8$ yr, and corresponding evaporation times
of $\sim 10^{10}$ yr (e.g., Subramaniam et al. \cite{Sub1993}), 
and may thus survive for one Hubble time.

\section{Conclusions}

We have analysed high-resolution Gemini/Hokupa'a adaptive optics 
and HST/NICMOS data of the Arches cluster near the Galactic Center
with respect to spatial variations in the mass function and their
implications for cluster formation. A detailed comparison of the Gemini data
to HST/NICMOS observations allows us to investigate the instrumental 
characteristics of PSF fitting photometry with the Hokupa'a AO system.

\subsection{Technical comparison of Gemini/Hokupa'a with HST/NICMOS}

The calibration of the Gemini/Hokupa'a data 
of the Arches cluster using HST/NICMOS data from Figer et al. (\cite{FKM}) 
allows us to carry out a detailed technical comparison of the two datasets. 
Maps of photometric residuals show a strong dependence of 
the calibration error on the stellar density within the field.
In particular, the vicinity of fainter objects to bright stars 
causes the Gemini magnitude to be underestimated in comparison
with HST/NICMOS. This is understandable as the uncompensated seeing
halos of bright stars enhance the background in a non-homogeneous 
way, thereby causing an overestimation of the background and a 
subsequent underestimation of the faint objects' magnitude.
Conversely, the flux of very bright sources seems to be overestimated.
The correlation of the photometric residual with the position
of bright stars is more pronounced in $K^\prime$, where crowding
is the dominant source of photometric uncertainty, while the 
effect of angular anisoplanatism is less severe. 
In the $H$-band the anisoplanatism is more pronounced, and hence 
photometric uncertainties are a blend of uncertainties due to the 
distance to the guide star and due to the proximity to bright sources.

The incompleteness of the luminosity function measured with Hokupa'a 
increases faster at fainter magnitudes than the incompleteness observed 
in the NICMOS LF despite the comparable detection limit in both datasets.
As expected, this effect is particularly pronounced in the dense
cluster center, where crowding is most severe.

As the Strehl ratio determines the amount of light scattered into
the seeing induced halo around each star, a good SR is crucial
to achieve not only a diffraction limited spatial resolution,
but to benefit from the adaptive optics correction in dense fields
containing a wide range of magnitudes. In the Arches dataset, the 
SR of only 2.5 \% in $H$ and 7 \% in $K^\prime$ as compared to 95 \% 
in F160W and 90 \% in F205W (NICMOS) is clearly the limiting 
factor for crowded field photometry. As Hokupa'a was initially designed 
and developed for the 3.6m CFH telescope, its performance at the 8m Gemini 
telescope is naturally constrained by the limited number of only 36 actuators.
In the case of the Arches science demonstration data, additional constraints
were given by the seeing, the high airmass due to the low latitude of the
Galactic Center, and the guide star magnitude. Under better observing 
conditions and with a brighter guide star Strehl ratios of up to 30 \% can
be achieved with Hokupa'a at Gemini. Higher order AO systems are currently
capable to produce SRs of up to 50\%.

The Gemini ground-based AO data are comparable to the 
HST/NICMOS data in the resolution of bright sources 
($K^\prime < 18$ mag), and in  a non-crowded field. They do reach their
limitations in the densest cluster area and in the case where faint stars
are located close to a bright object. 
Higher Strehl ratios would of course reduce this unequality.

\subsection{Photometric results}

A strong colour gradient is detected over the field of the Arches cluster,
revealing an increase in visual extinction of approximately 
$9 < \Delta A_V < 15$ mag when progressing outwards from the cluster 
center. The visual extinction is estimated from the Rieke \& Lebofsky
(\cite{RL85}) extinction law to be $A_V \sim 24$ mag in the cluster center,
increasing to a maximum of $33 < A_V < 39$ mag in the vicinity, 
in accordance with Cotera et al. (\cite{Cotera2000}), who found a maximum of 
$A_V = 37$ mag in the Arches field.
Within the central 5\arcsec\ radius, however, no colour gradient is observed.
This indicates that the cluster center has been stripped of the
remaining dust and gas either by strong stellar winds from massive
stars or by photo-evaporation, or both. Beyond 5\arcsec, a linear 
increase in extinction is observed, suggesting an increasing 
amount of dust with distance from the cluster center.
Photo-evaporation due to
the strong UV radiation of the 8 WN7/8-stars and more than 100 O-stars 
found in the cluster center is most probably responsible for 
dust dissolution.

The $m160-m205$, $m205$ (equivalent to $H-K$, $K$) colour-magnitude 
diagrams derived from the HST and Gemini data sets
both show a bent main sequence following this colour trend.
The main sequence straightens out when correcting for this colour variation.
A spatial analysis of the CMDs reveals the bulk of the bright stars on the
Arches field to be located in the cluster center.

\subsection{Mass functions}

Present-day mass functions have been derived from the CMDs after linear correction of 
the colour trend and corresponding change in extinction over the field,
and selection of a reasonable main sequence colour cut.
The integrated mass function derived from the Gemini photometry
displays a slope of $\Gamma = -0.8 \pm 0.15$ for $6\!<\!M\!<\!65\ M_\odot$, 
less steep than the Salpeter slope of $\Gamma = -1.35$.
This value agrees
with the slopes derived from the HST data in the same manner 
(Sect. \ref{mfsec}), and with the values presented in FKM within 
the uncertainties. When the magnitudes are not corrected for 
differential extinction, the slope of the MF is significantly 
flatter, $\Gamma \sim -0.5 \pm 0.2$. Particularly in young 
star forming regions, the effects of differential extinction 
are thus clearly non-negligible.

The analysis of the radial dependence of the mass function reveals a 
very flat IMF in the immediate cluster center with a slope close to zero.
The IMF slope seems to increase outwards with $\Gamma = -1.0 \pm 0.3$ 
for $5 < R < 10\arcsec$ and $\Gamma = -1.7 \pm 0.7$ beyond 
$R > 10\arcsec$. We have created cumulative functions for the stars
in each radial bin, and performed a KS test to derive the significance
level of the variance in the mass distributions. The probability for 
the observed distributions to originate in the same mass function
is below 1\% when comparing each two of the three radial bins analysed.

\subsection{Cluster dynamics}

The flat mass function in the Arches center is 
a strong indication for mass segregation. While the center seems to 
be dominated by high-mass stars, the cluster edges display the
standard behaviour of a young stellar population. 
A similar radial dependence of the mass function is observed in the 
young, compact cluster NGC\,3603 (Grebel et al. 2002, in prep.), 
located in a normal star forming environment in the Carina spiral arm. 
The fact that two out of three compact young clusters found in the 
Milky Way, which have been analysed in such 
detail to date, display a flat mass function slope in the core indicates
that such a behaviour might be typical for starburst clusters and is not
restricted to the extreme GC environment.

Mass segregation in a compact, massive cluster can either be caused
by an enhanced production efficiency of massive stars, or by dynamical
segregation during the cluster's evolution.
We roughly estimate the present-day relaxation time of Arches 
from mass function considerations to be about a few Myr ($\sim 20\,t_{\rm cross}$), 
and thus of the same order of magnitude as the cluster age of $\sim 2$ Myr. 
Again, a similar timescale has been found for NGC\,3603 as well 
(Grebel et al. 2002, in prep.). The comparison with N-body simulations suggests, however,
that the dynamical evolution of massive clusters close to the Galactic Center
whipes out the initial conditions within less than 1 Myr. We are therefore not 
able to distinguish between primordial and dynamical mass segregation.
Both effects are intertwined at the current state of 
cluster evolution, such that a detailed dynamical analysis is crucial
for a thorough understanding of the formation process. This analysis 
has to await deep, high resolution infrared spectroscopy to obtain
radial velocities and proper motions for a significant fraction of
stars belonging to the cluster population. 

\begin{acknowledgements} 

We are grateful to the Gemini Team for providing the science demonstration data,
as well as for all kinds of useful information and support on 
data analysis issues. In particular, we would like to thank
Dr. Fran\c{c}ois Rigaut, Dr. Jean-Ren\'{e} Roy and Dr. Mark Chun for their help 
with data reduction problems. 
We thank Dr. Michael Odenkirchen for infinite patience in helpful discussions.
We also thank our referee Simon Portegies Zwart for discussions and useful
comments which helped to improve the paper, in particular with respect to the 
mass function discussion and dynamical considerations.

The work presented here is 
based on observations obtained at the Gemini Observatory, which is operated by 
the Association of Universities for Research in Astronomy, Inc., under a cooperative
agreement with the NSF on behalf of the Gemini partnership: the National Science 
Foundation (United  States), the Particle Physics and Astronomy Research Council 
(United Kingdom), the National Research Council (Canada), CONICYT (Chile), 
the Australian Research Council (Australia), CNPq (Brazil) and CONICET (Argentina),
and based on observations made with the NASA/ESA Hubble Space Telescope, obtained at the 
Space Telescope Science Institute, which is operated by the Association of Universities
for Research in Astronomy, Inc., under NASA contract NAS 5-26555. These observations
are associated with proposal No. 7364.
\end{acknowledgements}

%\listofobjects

\end{document}